\documentclass[12pt,titlepage]{article}

\newif\ifdraft
\ifx\draftmode\undefined
  \draftfalse
\else
  \drafttrue
\fi

\usepackage{graphicx}
\DeclareGraphicsExtensions{.jpg,.eps,.pdf}

\ifdraft
  
\else
  \usepackage[nolists,nomarkers]{endfloat}
  \AtBeginDelayedFloats{}
\fi

\usepackage{helvet}

\usepackage{amssymb}

\usepackage{url}

\usepackage{enumitem}

\usepackage{cite}
\let\citeleft=(
\let\citeright=)

\usepackage{graphicx}
\usepackage{caption}
\usepackage{subcaption}

\usepackage{mathtools}

\usepackage{multirow}
\usepackage{tabularx}
\usepackage{array, booktabs}
\usepackage{makecell}
\usepackage[T1]{fontenc}

\newtagform{brackets}{[}{]}
\usetagform{brackets}

\graphicspath{{figs/}}
\usepackage{xcolor}
\usepackage{color, soul}
\usepackage{marginnote}
\usepackage[disable]{todonotes}

\begin{document}
\newcommand{\comment}[1]{\todo[inline]{#1}}

\newcommand\rightnote{\normalmarginpar\marginnote}
\newcommand\leftnote{\reversemarginpar\marginnote}
\newcommand\leftnotepar{\reversemarginpar\marginpar}
\newcommand\rightnotepar{\normalmarginpar\marginpar}
\newcommand\partextwidth{\parbox{1\textwidth}}
\newcommand\yellowbox{\colorbox{yellow}}
\newcommand\redtext{\color{red}}

\renewcommand\hl{\textnormal}
\renewcommand\yellowbox{\textnormal}
\renewcommand\partextwidth{\textnormal}
\renewcommand\leftnote{\comment}
\renewcommand\rightnote{\comment}
\renewcommand\rightnotepar{\comment}
\renewcommand\leftnotepar{\comment}

\bibliographystyle{mrm}

\begin{titlepage}

\begin{center}
\textbf{\Large A Regional Bolus Tracking and Real-time B$_1$ Calibration Method for Hyperpolarized $^{13}$C MRI}
\end{center}
\bigskip

\begin{center}
Shuyu Tang$^{1,2}$, Eugene Milshteyn$^{1,2}$, Galen Reed$^{3}$, Jeremy Gordon$^{2}$,  Robert Bok$^{2}$, \\
\yellowbox{ 
Xucheng Zhu$^{1,2}$, Zihan Zhu$^{1,2}$,
}
 Daniel B. Vigneron$^{1,2}$, Peder E.Z. Larson$^{1,2}$
\end{center}
\leftnote{\redtext{RA.1}}

\vspace*{0.1in}
\noindent
\noindent
$^1$UC Berkeley-UCSF Graduate Program in Bioengineering, University of California, San Francisco and University of California, Berkeley

\noindent
$^2$Department of Radiology and Biomedical Imaging,
University of California - San Francisco, San Francisco, California.  

\noindent
$^3$HeartVista, Los Altos, California, USA

\noindent
\noindent
{\em Address correspondence to:} \\
	Shuyu Tang \\
	Byers Hall, Room 102 \\
	1700 4th St \\
	San Francisco, CA \ 94158 \\
E-MAIL: shuyu.tang@ucsf.edu 

\noindent

\noindent
Approximate Word Count: \hl{246}\leftnotepar{\redtext{R2.1\newline RA.1}}  (abstract)  \hl{4793}  (body) \\

\noindent
Submitted Dec 22, 2017, to {\it Magnetic Resonance in Medicine} as a Full Paper.\\

\end{titlepage}

\section*{Abstract}
\setlength{\parindent}{0in}


Purpose: Acquisition timing and B$_1$ calibration are two key factors that affect the quality and accuracy of hyperpolarized $^{13}$C MRI. The goal of this project was to develop a new approach using regional bolus tracking to trigger Bloch-Siegert B$_1$ mapping and \hl{real-time B$_1$ calibration based on regional B$_1$ measurements}\leftnotepar{\redtext{R2.2}}, followed by dynamic imaging of hyperpolarized $^{13}C$ metabolites in vivo. 

Methods: 
The proposed approach was implemented on a system which allows real-time data processing and real-time control on the sequence. Real-time center frequency calibration upon the bolus arrival was also added. The feasibility of applying the proposed framework for in vivo hyperpolarized $^{13}$C imaging was tested on healthy rats, tumor-bearing mice and \hl{a healthy volunteer} \leftnotepar{\redtext{RA.1}} on a clinical 3T scanner following hyperpolarized [1-$^{13}$C]pyruvate injection. Multichannel receive coils were used in the human study. 

Results: Automatic acquisition timing based on \hl{either regional bolus peak or bolus arrival} \rightnotepar{\redtext{RA.1}} was achieved with the proposed framework. Reduced blurring artifacts in real-time reconstructed images were observed with real-time center frequency calibration. Real-time computed B$_1$ scaling factors agreed with real-time acquired B$_1$ maps. Flip angle correction using B$_1$ maps results in a more \hl{consistent} \leftnotepar{\redtext{R1.4}}  quantification of metabolic activity (i.e, pyruvate-to-lactate conversion, k$_{PL}$). Experiment recordings are provided to demonstrate the real-time actions during the experiment.

Conclusion: The proposed method was successfully demonstrated on animals and \hl{a human volunteer}\rightnotepar{\redtext{RA.1}} , and is anticipated to improve the efficient use of the hyperpolarized signal as well as the accuracy and robustness of hyperpolarized $^{13}$C imaging.   

\vspace{0.4in}
\setlength{\parindent}{0in}
{\bf Key words: Real-time, Bloch-Siegert, \textit{B$_1$} mapping, $^{13}$C, Hyperpolarized, MRI, metabolic imaging, Bolus tracking, Pyruvate}
\newpage

%
%
%
%
%
%


%
\section*{Introduction}

Magnetic resonance imaging with hyperpolarized $^{13}$C-labeled compounds prepared via dynamic nuclear polarization \cite{Ardenkjaer-Larsen:2003aa} has been used to non-invasively study metabolic processes in vivo\cite{Golman:2006aa}. \leftnotepar{\redtext{R1.7\newline R2.4\newline R2.20}} \hl{In this method, the hyperpolarized state of the agent and any metabolic products will irreversibly decay to thermal equilibrium due to metabolic conversion, RF depletion and T$_1$ relaxation.} Hyperpolarized [1-$^{13}$C]pyruvate has been widely used to monitor metabolic pathways \cite{Day:2007aa, Albers:2008aa,Schroeder:2008aa,Park:2010aa, Witney:2010aa,Rider:2013aa,Darpolor:2011aa, Kurhanewicz:2011aa,Nelson:2013aa,Cunningham:2016aa} and its feasiblity for human study has been demonstrated\cite{Nelson:2013aa,Cunningham:2016aa}. The T$_1$s of hyperpolarized [1-$^{13}$C]pyruvate and its downstream metabolites (e.g. lactate) at clinical field strengths are \hl{estimated to be 25s to 45s}\leftnotepar{\redtext{R1.8\newline R2.21}} \cite{Nelson:2013aa,Kazan:2013aa} in vivo. In light of the nonrenewable nature and fast decay of the hyperpolarized magnetization, acquisition timing and transmit power (B$_1$) calibration are two key factors that affects the quality and accuracy of hyperpolarized $^{13}$C imaging.

 \rightnote{\redtext{R2.3\newline R2.22}}\hl{Bolus tracking for triggering the acquisition has been used in clinical MRI for proton MR angiography}\cite{foo_automated_1997}. \hl{Appropriate acquisition timing} for hyperpolarized $^{13}$C imaging is useful in several aspects: a) Excitation before the bolus arrival may saturate the nonrecoverable hyperpolarized signal particularly near the coils conductive elements where the B$_1$ can be elevated. b) Many variable flip schemes \yellowbox{\cite{xing_optimal_2013,Nagashima:2008aa,zhao_gradient-echo_1996,Maidens:2016aa} }\leftnotepar{\redtext{R1.13}} that optimize the hyperpolarized MRI signal sampling require knowledge of the bolus arrival. c) It is more straightforward for kinetic modeling to start signal sampling after the bolus maximum \cite{durst_bolus_2014} to eliminate the need to account for the input function. d) Inconsistent acquisition timing leads to quantification errors when metabolite to substrate ratios are used as a quantitative metric \cite{yen_imaging_2009, durst_bolus_2014}. Currently, in vivo hyperpoliarized $^{13}$C imaging protocols typically start at a certain delay time after bolus injection. The delay time is determined based on prior knowledge, which can be unreliable due to the inherent physiological variability of each individual. This is particularly problematic in human subjects in which 12 s variations in bolus arrival have been observed \cite{Nelson:2013aa}. This variability can be further exacerbated in tumors where the vascularization and perfusion are highly inconsistent over subjects \cite{gillies_causes_1999}.  Recently, a bolus tracking method using a slab FID as tracking signal was demonstrated \cite{durst_bolus_2014} for automatic acquisition timing. However, this method didn't reflect the signal variation within the imaging slab. In this study, bolus tracking was improved by using the bolus signal from a region of interest (ROI) on a tracking image rather than from the entire excitation slab. 

In the context of hyperpoliarized $^{13}$C MRI, B$_1$  calibration is crucial for variable flip angle schemes \cite{xing_optimal_2013,Nagashima:2008aa, Maidens:2016aa} and quantification of metabolic activities \cite{Sun:2017aa}. Due to the virtually non-existent endogenous $^{13}$C signal, $^{13}$C B$_1$ calibration is typically performed on external phantoms, but this approach does not account for the variability of subject loading. Bloch-Siegert B$_1$ mapping has been applied to hyperpolarized $^{13}$C imaging \cite{lau_integrated_2012, schulte_transmit_2011} and is advantageous due to its short acquisition time and efficient use of the hyperpolarization signal. However, real-time power compensation has not yet been accomplished to calibrate the flip angle during the scan.
   
This work presents a new approach using regional bolus tracking to trigger Bloch-Siegert B$_1$ mapping and real-time RF power compensation based on \hl{regional B$_1$ measurements }\rightnote{\redtext{R2.2}} followed by dynamic imaging of hyperpolarized $^{13}$C metabolites. Real-time center frequency calibration upon bolus arrival was also implemented. Thermally polarized $^{13}$C phantom experiments were performed to validate Bloch-Siegert B$_1$ mapping. The feasibility of applying the proposed framework for in vivo hyperpolarized $^{13}C$ imaging was demonstrated on healthy rats, tumor-bearing mice and \hl{a healthy volunteer}  \leftnotepar{\redtext{RA.1}}  on a clinical 3T scanner. This proposed method was designed to improve the efficient use of the hyperpolarized signal as well as the accuracy and the robustness of hyperpolarized $^{13}$C imaging.

\section*{Methods}

\subsubsection*{Real-time Hyperpolarized $^{13}$C MRI}

\begin{figure}
\centering  
\includegraphics[width=1\textwidth]{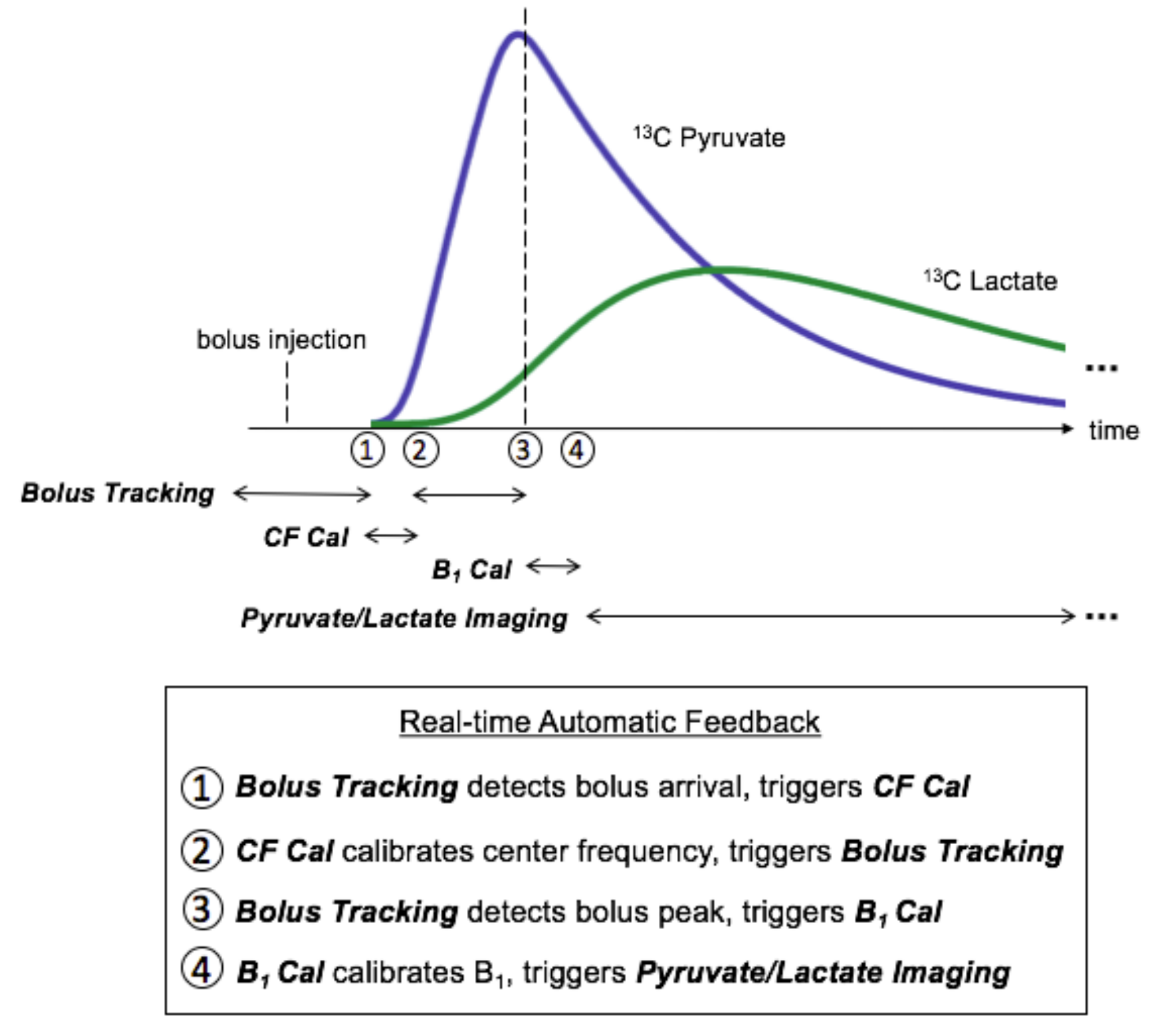}
 \caption
    {
Overview of the proposed scheme. Bolus tracking starts before the bolus injection.  Real-time center frequency calibration (`CF Cal') based on a slab FID is triggered upon ROI bolus arrival, while Bloch-Siegert B$_1$ mapping and real-time ROI B$_1$ calibration (`B$_1$ Cal') are triggered at the ROI bolus peak. The sequence triggered after B$_1$ calibration for most experiments in this study is alternate pyruvate/lactate dynamic imaging, and could be replaced by any hyperpolarized  $^{13}$C sequence for other studies. 
    }
  \label{fig:seq_scheme}
\end{figure}

The proposed scheme is illustrated in Fig. \ref{fig:seq_scheme}. \rightnotepar{\redtext{R2.5}}\hl{ROIs for both bolus tracking and B$_1$ calibration are prescribed according to proton anatomical images before starting $^{13}$C sequences.} The bolus tracking sequence starts before the hyperpolarized substrate is injected. Real-time center frequency calibration based on the acquired slab frequency spectrum can be triggered upon the bolus arrival. When the peak bolus signal in the prescribed ROI is detected, \rightnotepar{\redtext{R2.23}}\hl{Bloch-Siegert B$_1$ mapping is triggered and the RF power for all sequences is then calibrated in real time based on the measured ROI B$_1$.} \rightnotepar{\redtext{RA.1}}  \hl{The sequence triggered upon the completion of B$_1$ calibration for most experiments in this study is alternate pyruvate/lactate dynamic imaging, which could be replaced by any hyperpolarized $^{13}$C sequence for other studies.}  Our proposed scheme \hl{was implemented}\leftnotepar{\redtext{R2.24}} on a GE Signa MR 3T scanner (GE Healthcare, Waukesha, WI) using a commercial software (RTHawk, HeartVista, Los Altos, CA) which allows for real-time reconstruction of acquired data and feedback control \hl{of the pulse sequence}\rightnotepar{\redtext{R2.25}}. The software was installed on a workstation (2.4GHz, 16 proccessors, 64 GB RAM) running the Ubuntu operating system. 

\begin{figure}
\centering  
\includegraphics[width=1\textwidth]{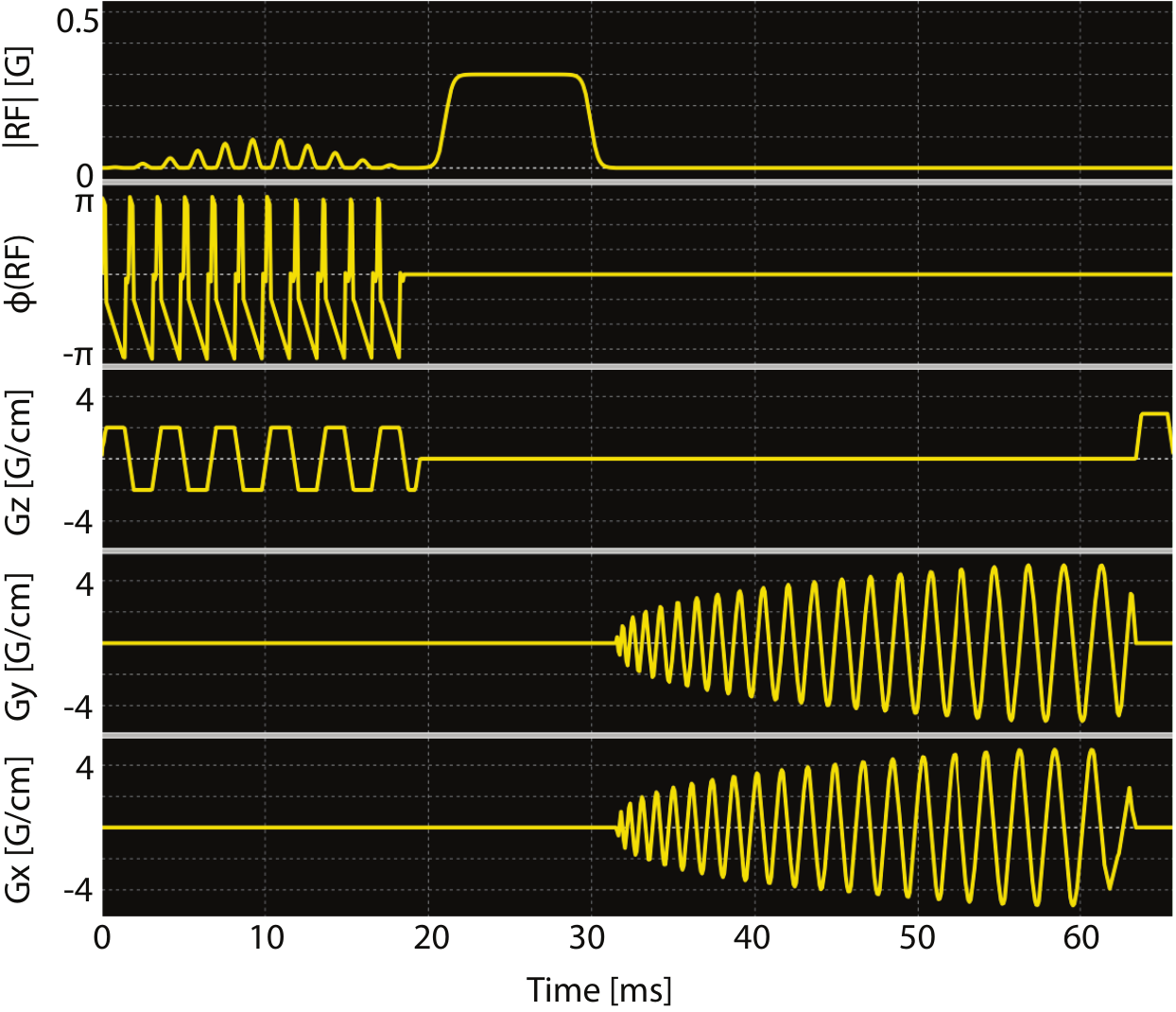}
  \caption
    {
    Metabolite-specific imaging sequence used in this study, where all instances included a singleband spectral-spatial excitation pulse (passband 120 Hz, stopband 600 Hz) and a single-shot spiral readout. The sequence shown also includes an off-resonance Fermi pulse (T$_{RF}$ = 12 ms, $\omega_{RF}$ = $\pm$4.5 kHz) for Bloch-Siegert B1 mapping, while bolus tracking and pyruvate/lactate dynamic imaging also used this sequence but without Fermi pulse and its associated delay. Other key parameters for bolus tracking \hl{were} FA 5$^o$ (pyruvate), TR 1s; and for Bloch-Siegert B1 mapping \hl{were} FA 10$^o$ (pyruvate), TR 200ms. 
    }
  \label{fig:seq_waveform}
\leftnote{\redtext{R2.35}}   
\end{figure}

The bolus tracking sequence used a singleband spectral-spatial excitation pulse and a single-shot spiral readout, similar to Fig. \ref{fig:seq_waveform} but without the Fermi pulse and associated delay. In these studies, this was used for selective imaging of [1-$^{13}$C]pyruvate, but could be adapted for other metabolites (e.g. $^{13}$C-urea). The algorithm for tracking the maximum bolus signal was implemented based on prior works \cite{foo_automated_1997, durst_bolus_2014}. The bolus signal is the mean value of a prescribed ROI on the tracking image. \leftnotepar{\redtext{R2.6}}\hl{Our bolus tracking acquisition consists of two modules, i.e., noise calibration and bolus tracking, which can be performed independently. Noise calibration is used to determine the tracking threshold. If the tracking threshold has been previously computed, bolus tracking can be performed without running noise calibration.} In the noise calibration module, the tracking threshold $S_{thr}$ is determined as $S_{thr} = c_{thr} \sigma =c_{thr} \frac{m_{noise}}{\sqrt{\frac{\pi}{2}}}$, where $\sigma$ is the standard deviation of Gaussian noise in the complex image, $m_{noise}$ is the mean value of magnitude images of noise and $c_{thr}$ is a scaling factor. 100 calibration scans (TR = 200 ms) are performed to compute $m_{noise}$ which is converted to $\sigma$ \cite{Gudbjartsson:1995aa}. Given the noise distribution of the background in magnitude images follows a Rayleigh distribution\cite{Gudbjartsson:1995aa}, if a probability $P_{thr}$ that a noise signal is above the tracking threshold is required, the proportional relationship between $S_{thr}$ and $\sigma$ can be derived using the cumulative distribution function of a Rayleigh distribution. For example, $c_{thr}$ = 3 would result in a $P_{thr}$ of 0.001\%. In the bolus tracking module, the sequence is resumed automatically if the bolus peak is not detected. When a cumulative number $n_{cum}$ of signal increases is detected and all these signals are above the tracking threshold as well as greater than $u_{peak}$ fraction of its previous signal, the program starts to update the peak value. When the bolus signal is lower than $u_{peak}$ fraction of the current peak signal, detection of bolus peak will be reported. \leftnotepar{\redtext{R2.7}} \hl{In our study, $u_{peak}$ was determined based on estimated hypolarization loss due to T$_1$. For example, under the assumptions of a T$_1$ of 30s, an excitation flip angle of 5$^{o}$, no more pyruvate from the bolus coming into the ROI and neglecting metabolic conversion, the hyperpolarization loss over a TR of 1s for the bolus tracking sequence would be 1 - cos(5$^{o}$ $\times$ $\pi$/180) $\times$ exp(-1/30) = $\sim$0.04, and the remaining hyperpolarization ($\sim$0.96) was used as $u_{peak}$.} The use of cumulative increase rather than successive increase as used in the previous work \cite{durst_bolus_2014} aims to improve the robustness of the tracking algorithm to motion and injection with unstable rates. For some experiments, during the time that bolus signal was below the current signal peak but above $u_{peak}$ threshold, a short repetition time was used to shorten the time interval between actual bolus peak and detected bolus peak.

The Bloch-Siegert B$_1$ mapping sequence\cite{sacolick_b_2010, lau_integrated_2012, schulte_transmit_2011} (Fig. \ref{fig:seq_waveform}) shared the similar excitation and readout as bolus tracking, using a singleband spectral-spatial excitation pulse and a single-shot spiral readout, whereas an off-resonance Fermi pulse ($K_{BS}=6.76 rad/G^2$ ) was inserted in between. The phase difference between two Bloch-Siegert B$_1$ mapping sequences with opposite $\omega_{RF}$ was used to calculate B$_1$. Taking the phase difference removes receiver phase and minimizes the influence of B$_0$ off-resonance frequency \cite{sacolick_b_2010}. Parameters used in this study were designed to measure a maximum B$_1$ of 0.48G. The two phase maps were masked based on corresponding magnitude images in order to eliminate noisy phases due to low signal. The threshold is determined as S$_{B_1} = c_{B_1}\sigma$, where c$_{B_1}$ is a scaling factor and noise standard deviation $\sigma$ was obtained from the bolus tracking sequence. The ratio of the desired B$_1$ to the measured ROI B$_1$ was passed to all sequences as a scaling factor to calibrate transmit power in real time.

\hl{The center frequency} \leftnotepar{\redtext{R2.26}}  calibration sequence consists of a sinc excitation pulse with slab selection gradient and a 102 ms readout with a 5kHz bandwidth. The acquired spectrum is used to calibrate center frequency for all sequences in real time. Alternate pyruvate and lactate dynamic imaging shared the same excitation and readout as bolus tracking, using a singleband spectral-spatial excitation pulse and a single-shot spiral readout, but with alternating excitation frequencies and different flip angles to image pyruvate and lactate.

All methods described are available through the HeartVista research collaboration portal (\verb|https://www.heartvista.com/git/shuyu/hv_research_shuyu|).

\subsubsection*{Phantom and Animal Experiments}
\leftnote{\redtext{RA.1}}

Validations of the Bloch-Siegert B$_1$ mapping sequence were performed on a $^{13}$C/$^{1}$H birdcage coil as well as on a $^{13}$C figure-8 transceiver coil. For experiments using the $^{13}$C/$^{1}$H birdcage coil, a cylindrical ethylene glycol phantom with a base of 3.6cm diameter was used. Axial $^{13}$C B$_1$ maps at four transmit powers (50\%, 75\%, 100\% and 150\% of the calibrated power) were acquired  (FA $90^{o}$, slice thickness 10mm, Fermi pulse duration $T_{RF}$ 12ms, frequency offset $\omega_{RF}$ $\pm4.5 kHz$, FOV 10cm, in-plane resolution 2.5 $\times$ 2.5mm, TR 2s, NEX 100). For experiments using the $^{13}$C figure-8 transceiver coil, a cup (cross-section diameter from 5.5 to 7.5cm, height 7.5cm) filled with oil was placed on top of the coil. An axial $^{13}$C magnitude map with a nominal $180^o$ excitation (FA $180^{o}$, slice thickness 100mm, FOV 20cm, in-plane resolution 5 $\times$ 5mm, TR 1s, NEX 100) and its corresponding B$_1$ map (FA $60^{o}$, slice thickness 100mm, Fermi pulse duration $T_{RF}$ 8ms, frequency offset $\omega_{RF}$ $\pm3kHz$, FOV 20cm, in-plane resolution 5 $\times$ 5mm, TR 1s and NEX 200) were acquired. A 2D phase unwrapping algorithm \cite{TOWERS1991239} based on minimum spanning tree was implemented to correct phase images of B$_1$ mapping off line. 

Hyperpolarized $^{13}$C animal experiments were performed to test our proposed scheme using normal Sprague-Dawley rats and transgenic adenocarcinoma of mouse prostate (TRAMP) mice with $^{13}$C/$^{1}$H birdcage coils (8cm diameter for rats, 5cm diameter for mice). Two experiments were also performed on a rat with a $^{13}$C figure-8 transceiver coil. \hl{A total of six rats and two TRAMP mice were used to test the proposed scheme.}\leftnotepar{\redtext{R1.5}} All animal studies were conducted under protocols approved by the University of California San Francisco Institutional Animal Care and Use Committee (IACUC). Both rats and mice were anesthetized with isoflurane (1-2\%) and placed in a supine position on a heated pad throughout the duration of the experiments. [1-$^{13}$C]pyruvic acid (Sigma Aldrich, St. Louis, MO) mixed with 15mM trityl radical (GE Healthcare, Waukesha, WI) and 1.5mM Gd-DOTA (Guerbet, Roissy, France) was polarized in a HyperSense dissolution DNP system (Oxford Instruments, Abingdon, UK) at 1.35K and 3.35T for $\sim$1h. A 4.5mL volume of 80mM NaOH and 40mM Tris buffer was used as dissolution media, resulting in a 80mM [1-$^{13}$C]pyruvate solution, with final pH of 6-8. The hyperpolarized [1-$^{13}$C]pyruvate was injected into the animal via tail vein catheters, $\sim$2.6mL for each rat and $\sim$350$\mu$L for the mouse. Each animal received two injections.

All animal experiments were performed using the proposed scheme (Fig. \ref{fig:seq_scheme}). $^{13}$C sequence parameters for animal experiments are shown in Table \ref{tab:seq_params}. Some experiments were performed with real-time center frequency calibration. The desired B$_1$ of the Fermi pulse in B$_1$ mapping sequence for all experiments was always set to 0.3G (344\% of B$_1$ required for a $90^{o}$ excitation of the spectral-spatial pulse used in this study). Using the measured transmit B$_1$ maps, all in vivo results \hl{were compensated} \leftnotepar{\redtext{R2.27}}for spatial B$_1$ variations and flip angles between sequences. RF power measurements are shown as relative B$_1$ maps which is the measured B$_1$ normalized by the desired B$_1$. For some experiments, data from alternate pyruvate and lactate dynamic imaging was used to quantify pyruvate-to-lactate conversion rate (k$_{PL}$) based on a two-site exchange model \cite{harrison_comparison_2012} using non-linear least-squares fitting. 

For rat experiments using the $^{13}$C/$^{1}$H birdcage coil, an anatomical localizer was acquired using proton 3D bSSFP sequence (FOV 16 $\times$ 8 $\times$ 4.8cm, Matrix size 256 $\times$ 256 $\times$ 76). A total number of five injections were performed on three rats using the proposed scheme with different combinations of injection times (8s, 12s) and transmit gains (100\%, \hl{120\%} \leftnotepar{\redtext{R2.28}} of pre-scan power calibration). Real-time center frequency calibration were not performed in these studies. A $^{13}$C urea phantom was used for frequency and pre-scan power calibrations. $^{13}$C images were acquired on the axial kidney plane. ROIs for bolus tracking and B$_1$ calibration were both placed on the left kidney.

For rat experiments using the $^{13}$C transceiver coil, an anatomical localizer was acquired using proton T2-weighted fast spin echo sequence (FOV 6 $\times$ 6cm, Matrix size 256 $\times$ 256) with the scanner body coil. The rat was positioned in a way that right kidney was about 1 cm further than left kidney from the bottom $^{13}$C transceiver coil. This set up was designed to produce a distinct B$_1$ variation between two kidneys. Two injections were performed using the proposed scheme with the same parameters but with different tracking/B$_1$ calibrating ROIs, one on right kidney and the other on left kidney.  A $^{13}$C urea phantom embedded on the coil was used for pre-scan frequency calibration. Real-time center frequency calibration was performed at the bolus arrival.  $^{13}$C images were acquired on the axial kidney planes with an injection time of 10s and transmit gain set based on pre-scan with the $^{13}$C-urea phantom. 

For the TRAMP experiment, an anatomical localizer was acquired using proton T2-weighted fast spin echo sequence (FOV 6 $\times$ 6cm, Matrix size 256 $\times$ 256). Real-time center frequency calibration was performed at the bolus arrival. $^{13}$C images were acquired on the axial tumor plane with a injection time of 12s and transmit gain the same as the calibrated power in pre-scan. A $^{13}$C urea phantom was used for pre-scan frequency and power calibration. ROIs for bolus tracking and B$_1$ calibration were both in the tumor. 

\subsubsection*{Volunteer Study}

\leftnotepar{\redtext{RA.1}} \yellowbox{\partextwidth{One human study was performed to test the proposed scheme (Fig. \ref{fig:seq_scheme}). A healthy volunteer was recruited with institutional review board approval and provided with written informed consent for participation in the study. The volunteer was a 37-year-old male. An Investigational New Drug approval was obtained from the U.S. Food and Drug Administration for generating the agent and implementing the clinical protocol. 1.47g of Good Manufacturing Practices (GMP) [1-$^{13}$C]pyruvate (Sigma Aldrich, St. Louis, MO) mixed with 15mM electron paramagnetic agent (EPA) (AH111501, GE Healthcare, Oslo, Norway) was polarized using a 5T SPINlab polarizer (General Electric, Niskayuna, NY) for 3h before being rapidly dissolved with 130$^o$C water and forced through a filter that removed EPA. The solution was then collected in a receiver vessel and neutralized with NaOH and Tris buffer. The receive assembly that accommodates quality-control processes provided rapid measurements of pH, pyruvate and EPA concentrations, polarization, and temperature. In parallel, the hyperpolarized solution was pulled into a syringe (Medrad Inc, Warrendale, PA) through a 0.5$\mu$m sterile filter (ZenPure, Manassas, VA) and transported into the scanner for injection. The integrity of this filter was tested in agreement with manufacturer specifications prior to injection. A 0.43mL/kg dose of $\sim$250mM pyruvate was injected at a rate of 5mL/s via an intra-venous catheter placed in the antecubital vein, followed by a 20mL saline flush.}} 

\yellowbox{\partextwidth{
In this study, $^{13}$C axial brain images were acquired with in-house built birdcage transmit coil and 32-channel receive array \cite{mareyam_2017}. Different from animal experiments, excitation pulses for bolus tracking, B$_1$ mapping and metabolic-specific dynamic imaging were replaced with a different singleband spectral-spatial RF pulse (130Hz FWHM passband, 870Hz stopband) \cite{Gordon:2018aa}. Single-slice real-time B$_1$ calibration was triggered right after real-time center frequency calibration rather than on bolus peak. The ROI for both bolus tracking and B$_1$ calibration was on the brain tissue near the superior sagittal sinus. Following B$_1$ calibration, multi-slice 2D acquisitions were performed to cover the entire brain, and [1-$^{13}$C ]pyruvate, [1-$^{13}$C ]lactate and [1-$^{13}$C ]bicarbonate signals were acquired alternately. A $^{13}$C urea phantom embedded on the coil was used for pre-scan frequency calibration. Pre-scan power calibration was performed on a $^{13}$C ethylene glycol head phantom. Proton anatomical reference was acquired with body coil built in the scanner, using 3D T1-weighted spoiled gradient echo with FOV 26.6x26.6x16cm$^3$, matrix size 256x256x160. $^{13}$C sequence parameters for this study are presented in Table \ref{tab:human_seq_params}. 
\newline\newline
For the online reconstruction, multichannel k-space data were combined using sum of squares. In the offline processing, coil combination was performed by using pyruvate signals as coil sensivitiy maps \cite{zhu_2018}.
}} 
\leftnotepar{\redtext{RA.1}}

\section*{Results}

\begin{figure}
\centering  
\rightnote{\redtext{R2.36}}   
  \includegraphics[width=1\textwidth]{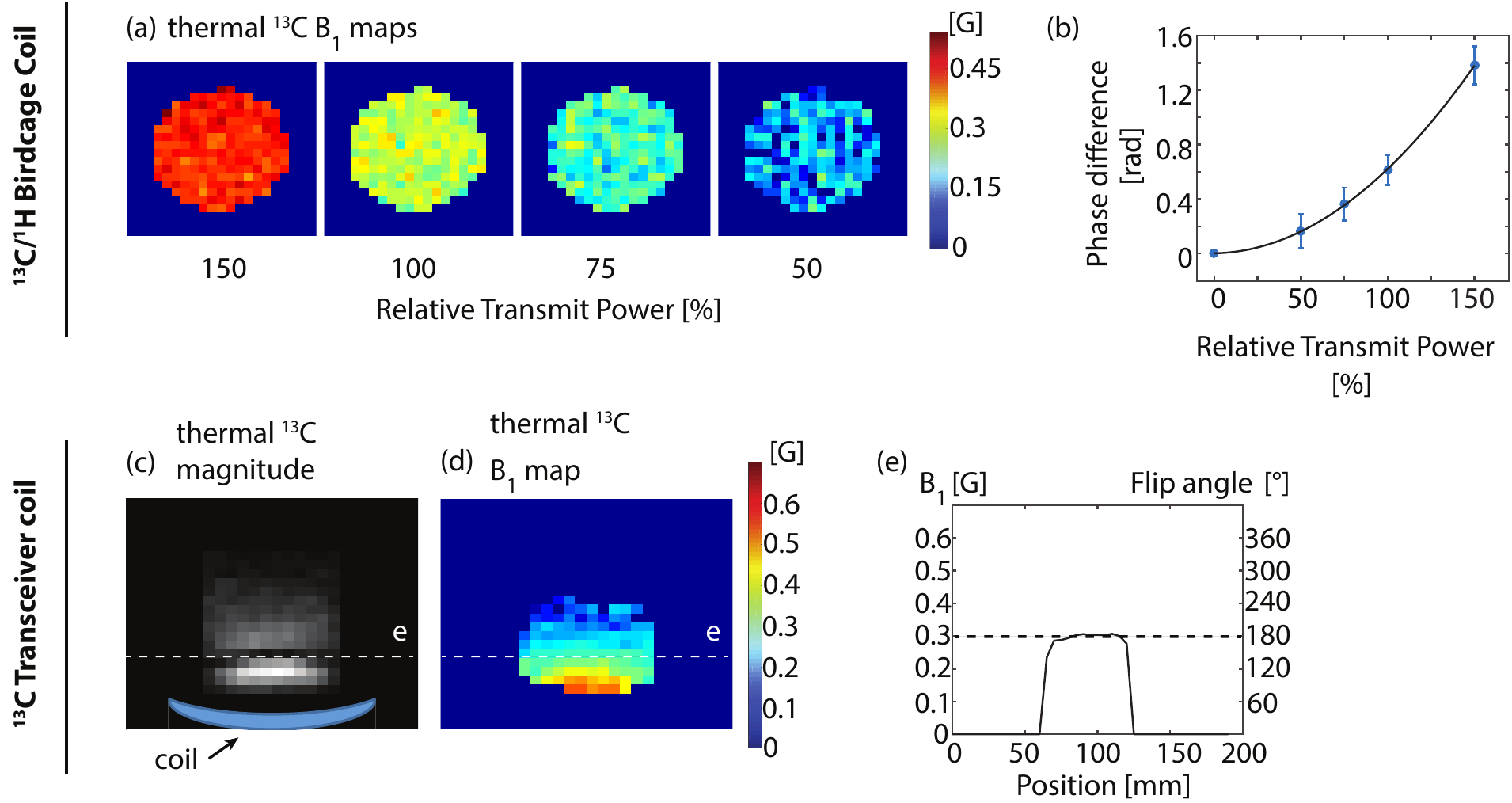}
  \caption
    {
    Validation of Bloch-Siegert B$_1$ mapping on the $^{13}$C/$^{1}$H birdcage coil using a cylindrical ethylene glycol phantom (a-b) and on the $^{13}$C figure-8 transceiver coil using a cup filled with oil (c-e). For the latter experiment, the oil cup was placed on the top of the $^{13}$C coil. In both experiments, the desired B$_1$ of the Bloch-Siegert pulse was 0.3G, at which the excitation pulse would produce the theoretically accurate flip angle. (a) Axial $^{13}$C B$_1$ maps of birdcarge coil with 50\%, 75\%, 100\% and 150\% relative to the calibrated transmit power in pre-scan. (b) A plot of Bloch-Siegert phase difference versus relative transmit power. Each data point corresponds to the mean value of the phantom area of each B$_1$ image in (a). The data point at the relative power of zero is estimated to be zero. The quadratic curve is computed by a least-squares fitting. (c) Axial $^{13}$C image of transceiver coil with a nominal 180$^{o}$ flip angle. The dark band in the image corresponds to a 180$^{o}$ signal null. (d) Corresponding axial B$_1$ map of (c), where the dark band corresponds to $\sim$0.3 G as expected. (e) A plot of B$_1$ value across the dark band and corresponding flip angles calculated based on the B$_1$ value.
    }
  \label{fig:phantom}
\end{figure}

Axial thermal $^{13}$C B$_1$ maps of the $^{13}$C/$^{1}$H birdcarge coil at 50\%, 75\%, 100\% and 150\% of the calibrated transmit power are shown in Fig. \ref{fig:phantom}a. The mean B$_1$ value of the phantom region versus the relative transmit power is plotted (Fig. \ref{fig:phantom}b) to demonstrate the quadratic relationship between Bloch-Siegert phase difference and transmit power. A comparison between the magnitude image of nominal $180^{o}$ excitation and its corresponding B$_1$ map of a thermal $^{13}$C phantom on the $^{13}$C transceiver coil is shown in Fig. \ref{fig:phantom}c and Fig. \ref{fig:phantom}d. The B$_1$ value along the dark band in Fig. \ref{fig:phantom}c and its corresponding theoretical excitation flip angle (0.3 G) is plotted in Fig. \ref{fig:phantom}e, demonstrating the accuracy of Bloch-Siegert B$_1$ mapping.

\begin{figure}
\leftnote{\redtext{R2.16}}   
\centering  
\includegraphics[width=1\textwidth]{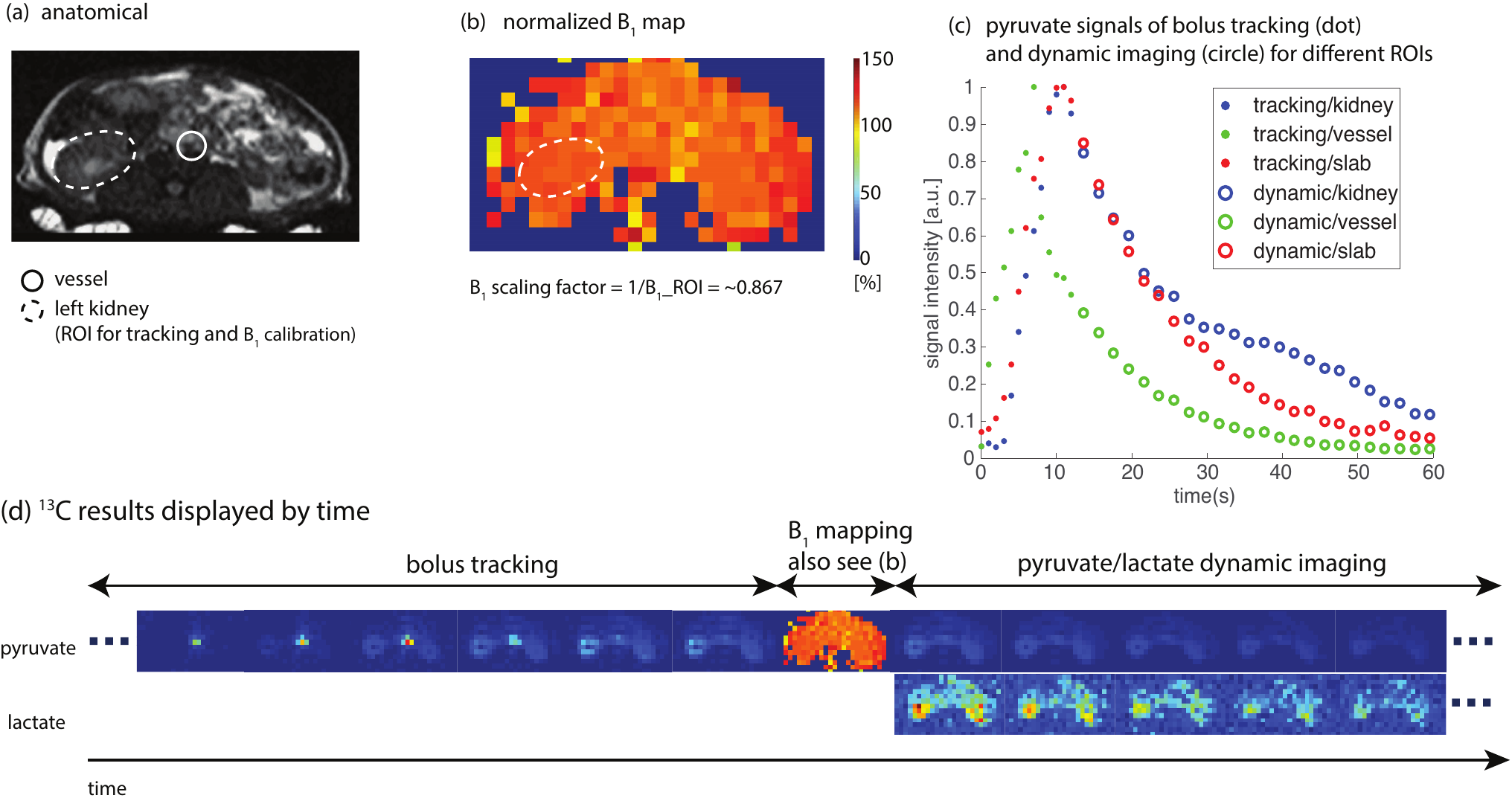}
  \caption
    {
Results of a hyperpolarized [1-$^{13}$C]pyruvate study in a normal rat using the proposed method (Fig. \ref{fig:seq_scheme}) with a birdcage coil. Real-time center frequency calibration was not performed in this study. The ROI for both bolus tracking and B$_1$ calibration was on the left kidney. Injection time was 8 s and initial power was purposely set to 120\% of the calibrated power in pre-scan. Sequence parameters are presented in Table \ref{tab:seq_params}. (a) Proton localizer. (b) Normalized $^{13}$C B$_1$ map. Real-time B$_1$ scaling factor ($\sim$0.87) matched up with the initial transmit power (120\%). (c) Normalized pyruvate signal curves in different ROIs. The ROI (left kidney) bolus peak was successfully detected. (d) $^{13}$C results displayed in the order of time. Every other timeframe is shown. The full set of images can be found in Sup. Fig. S1. Experiment recording: https://youtu.be/CN3mIrzmBT8.
    }
  \label{fig:healthy_rat}
\end{figure}

Results of a representative hyperpolarized [1-$^{13}$C]pyruvate experiment using the proposed scheme to image the kidneys of a healthy rat with $^{13}$C/$^{1}$H birdcage coil are shown in Fig. \ref{fig:healthy_rat}. Real-time center frequency calibration was not performed in this study. Fig. \ref{fig:healthy_rat}d \hl{displays} \leftnotepar{\redtext{R2.29}} every other timeframe of the real-time reconstructed data where magnitude images are normalized by the peak value of the corresponding metabolic series. The full set of images can be found in Sup. Fig. S1. The acquired B$_1$ map (Fig. \ref{fig:healthy_rat}b) was homogenous, as expected for this coil. \leftnotepar{\redtext{R2.10\newline R2.30}}\hl{The initial transmit power was purposely set to 120\% of the power calibrated on a thermal $^{13}$C phantom and resulted in a nominal B$_1$ scaling factor of 0.83 (1/1.2), which was in reasonable agreement with the real-time computed B$_1$ scaling factor of $\sim$0.87.} Normalized pyruvate signal curves combining the data of bolus tracking and pyruvate/lactate dynamic imaging are shown in Fig. \ref{fig:healthy_rat}c for different ROIs. These pyruvate signal curves demonstrate that the ROI (left kidney) bolus peak was successfully detected, and acquisition timing would be different if the tracking ROI was on the major vessels. 

\begin{figure}
\centering  
\leftnote{\redtext{R2.16}}   
  \includegraphics[width=1\textwidth]{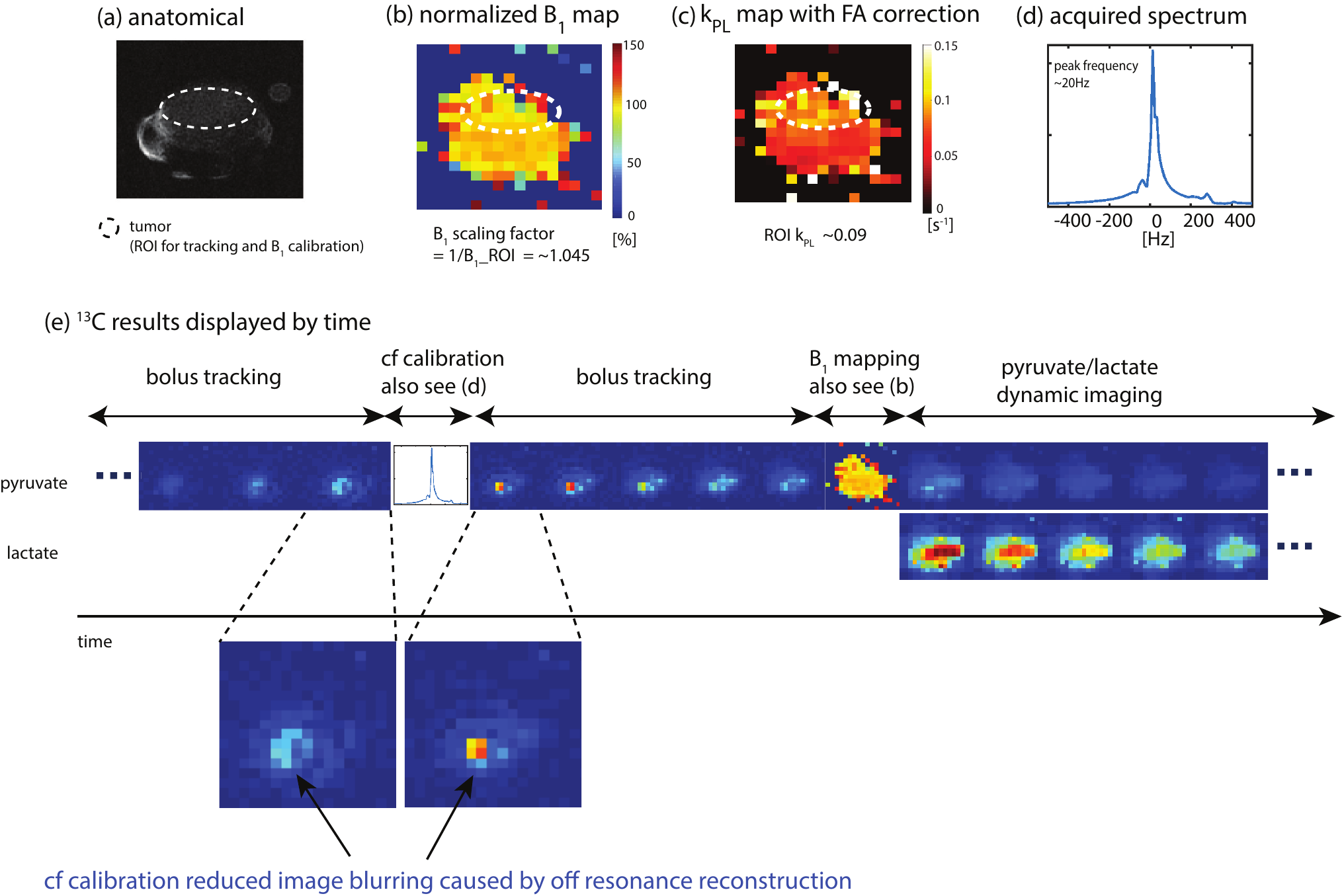}
\leftnote{\redtext{R2.11}}   
  \caption
    {
Results of a hyperpolarized [1-$^{13}$C]pyruvate study in a TRAMP mouse using the proposed method (Fig. \ref{fig:seq_scheme}) with a birdcage coil. The ROI for both bolus tracking and B$_1$ calibration was the tumor. Injection time was 12 s and initial power was the same as the calibrated power in pre-scan. Sequence parameters are presented in Table \ref{tab:seq_params}. (a) Proton localizer. (b) Normalized $^{13}$C B$_1$ map. (c) Pyruvate-to-lactate conversion rate($k_{PL}$) map with corrected flip angle. (d) Acquired frequency spectrum for center frequency calibration. ``Center frequency'' is abbreviated as ``cf'' in the figure. (e) $^{13}$C results displayed in the order of time. Every other timeframe is shown. The full set of images can be found in Sup. Fig. S2. Bolus tracking images before and after real-time center frequency calibration demonstrate that real-time center frequency calibration reduced off-resonance artifacts in real-time reconstructed images. Experiment recording: https://youtu.be/ViTDb3PzK3U.     
    }
  \label{fig:tramp}
\end{figure}

Results of a hyperpolarized [1-$^{13}$C]pyruvate experiment using the proposed scheme to image the tumor of a TRAMP mouse with $^{13}$C/$^{1}$H birdcage coil is shown in Fig. \ref{fig:tramp}. Fig. \ref{fig:tramp}e \leftnotepar{\redtext{R2.31}}\hl{displays} every other timeframe of the real-time reconstructed data where magnitude images are normalized by the peak value of the corresponding metabolic series. The full set of images can be found in Sup. Fig. S2. The frequency spectrum (Fig. \ref{fig:tramp}d) acquired from the imaging slab at the bolus arrival shows that the measured pyruvate frequency was 20Hz downfield from the frequency calibrated based on a thermal $^{13}$C phantom. Bolus tracking images right before and after real-time center frequency calibration in Fig. \ref{fig:tramp}e demonstrate that real-time center frequency calibration \leftnotepar{\redtext{R2.11}}\hl{reduced  blurring caused by off resonance reconstruction in real-time reconstructed images.} Fig. \ref{fig:tramp}b depicts the normalized B$_1$ map which is homogenous as expected for the coil. The real-time B$_1$ scaling factor (Fig. \ref{fig:tramp}b) was ~1.05, indicating a 5\% difference from the power calibrated on a thermal $^{13}$C phantom. 
Tumor k$_{PL}$ (Fig. \ref{fig:tramp}c) fitted using pyruvate and lactate signals was $\sim$ 0.09 s$^{-1}$, agreeing with prior works\cite{chen_2017} which showed a k$_{PL}$ range of 0.03 to 0.08 s$^{-1}$ for high grade TRAMP tumor.

\begin{figure}
  \leftnote{\redtext{RA.2\newline R2.16}}   
\centering  
  \includegraphics[width=1\textwidth]{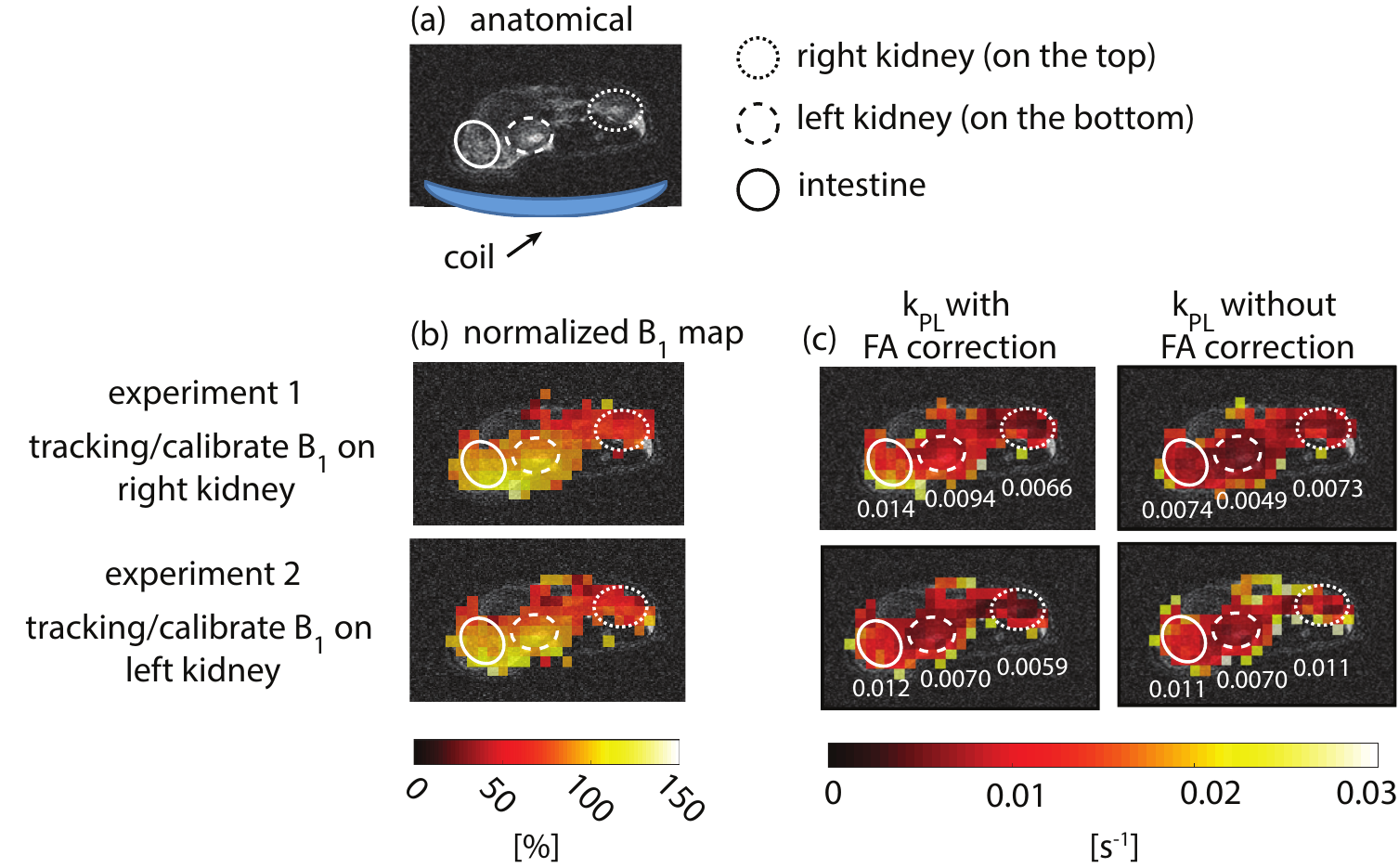}
   
  \caption
    {
Results of two hyperpolarized [1-$^{13}$C]pyruvate studies in a normal rat using the proposed method (Fig. \ref{fig:seq_scheme}) with a $^{13}$C surface transceiver coil. The two experiments were performed with the same parameters except for different tracking/calibrating ROIs: experiment \#1 on right kidney and \#2 was on left kidney, where the left kidney is closer to the coil. Injection time was 10 s and initial power was the same as the calibrated power in pre-scan. Sequence parameters are presented in Table \ref{tab:seq_params}. (a) Proton localizer. (b) Normalized $^{13}$C B$_1$ maps. B$_1$ maps acquired in two experiments are consistent. (c) Estimated k$_{PL}$ with and without flip angle correction based on measured B$_1$ map. $k_{PL}$ values for left kidney, right kidney and intestine are \hl{labeled in the maps.} Using acquired B$_1$ maps to correct flip angle results in more consistent k$_{PL}$ estimations of those ROIs between the two experiments, demonstrating the importance of flip angle correction for k$_{PL}$ measurements. Experiment recordings: https://youtu.be/Mu3NW7Kog9M, https://youtu.be/fL5gVkPDw2o.
    }
    \leftnote{\redtext{R2.17}}
  \label{fig:liver_coil}

\end{figure}

Results of two hyperpolarized [1-$^{13}$C]pyruvate experiments using the proposed scheme to image the kidneys of healthy rats with one-sided $^{13}$C surface transceiver coil are shown in Fig. \ref{fig:liver_coil}. Two experiments were performed with the same parameters but with the different tracking/calibrating ROIs, one on the right and the other on the left kidney, where the left kidney was closer to the coil. B$_1$ maps (Fig. \ref{fig:liver_coil}b) acquired in two experiments are consistent and show that the left kidney experienced a 40\% higher B$_1$ than the right kidney did. This agreed with the real-time B$_1$ scaling factors, 1.49 and 1.07 for the tracking experiments on the left and right kidney, respectively. k$_{PL}$ maps fitted with and without flip angle correction for the two experiments are shown in Fig. \ref{fig:liver_coil}c, where k$_{PL}$ values of right kidney, left kidney and intestine are labeled. \rightnotepar{\redtext{R1.4\newline R1.11}}  \hl{Root mean squared errors of k$_{PL}$ values in the three labeled ROIs between two experiments are 0.0033 with flip angle correction and 0.0045 without flip angle correction, demonstrating that using acquired B$_1$ maps to correct flip angle results in more consistent k$_{PL}$ estimations.}

\begin{figure}
 \leftnote{\redtext{RA.1}}
  \yellowbox{\partextwidth{
\centering  
  \includegraphics[width=1\textwidth]{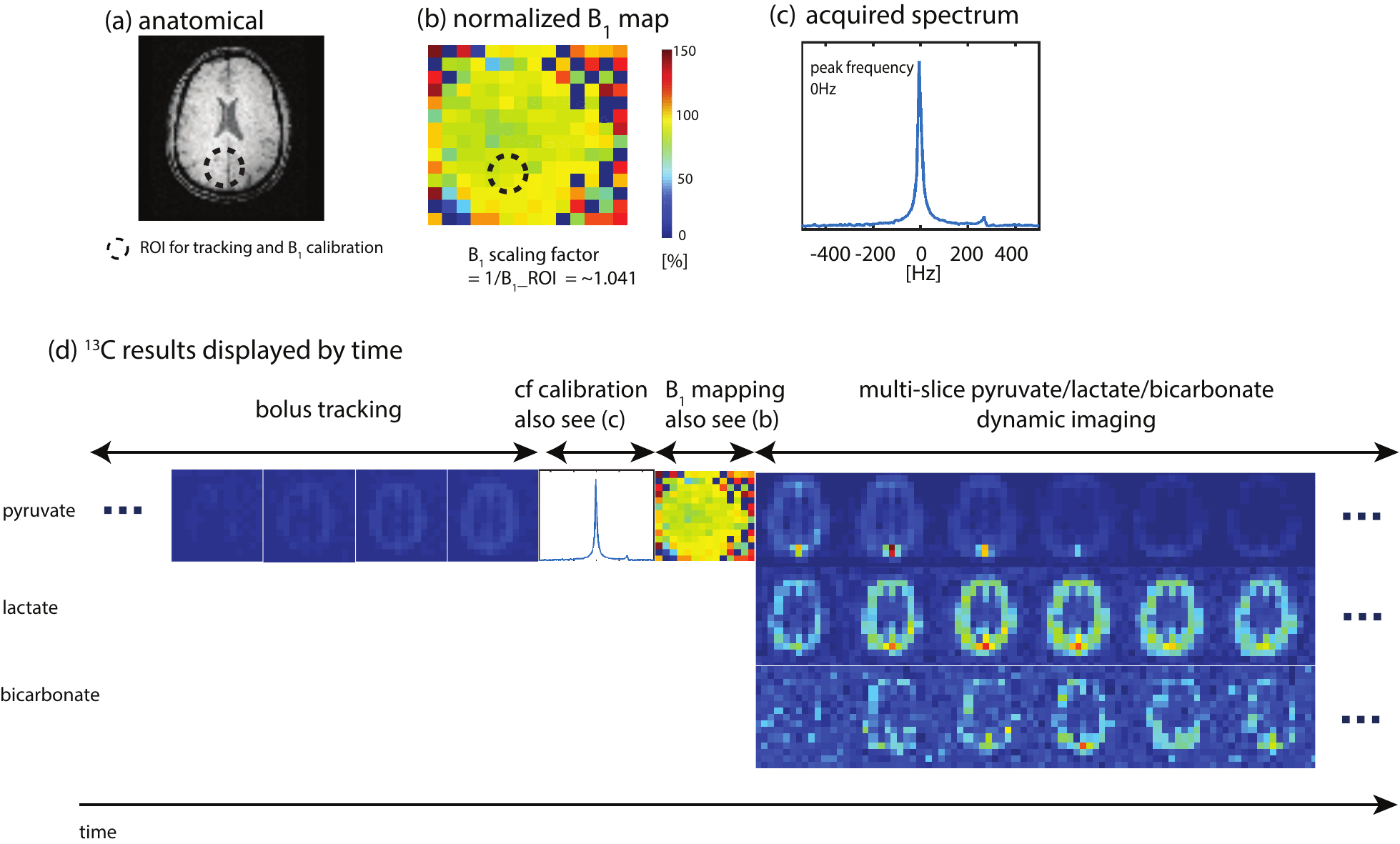}
    \caption
    {
Results of hyperpolarized [1-$^{13}$C]pyruvate studies on a healthy human volunteer using the proposed method (Fig. \ref{fig:seq_scheme}) with in-house built birdcage transmit coil and 32-channel receive array. In this study, single-slice real-time B$_1$ calibration was triggered right after real-time center frequency calibration. The ROI for both bolus tracking and B$_1$ calibration was on the brain tissue near the superior sagittal sinus of slice 5 (see Fig. 8 for slice 1-8). A multi-slice 2D acquisition was used for pyruvate/lactate/bicarbonate dynamic imaging. Initial power was the same as the calibrated power in pre-scan. Sequence parameters are presented in Table \ref{tab:human_seq_params}. (a) Proton image of slice 5. (b) Normalized $^{13}$C B$_1$ maps. (c) Acquired frequency spectrum for center frequency calibration. (d) $^{13}$C results of slice 5 displayed in the order of time. Multi-slice dynamic images can be found in Sup. Fig. S3. Sum-over-time images can be found in Fig. 8. Experiment recordings: https://youtu.be/Oq36Z7ayQ0g.
    }  
 \label{fig:human}
 }}
\end{figure}

\begin{figure}
 \leftnote{\redtext{RA.1}}
    \yellowbox{\partextwidth{
\centering  
\includegraphics[width=6in]{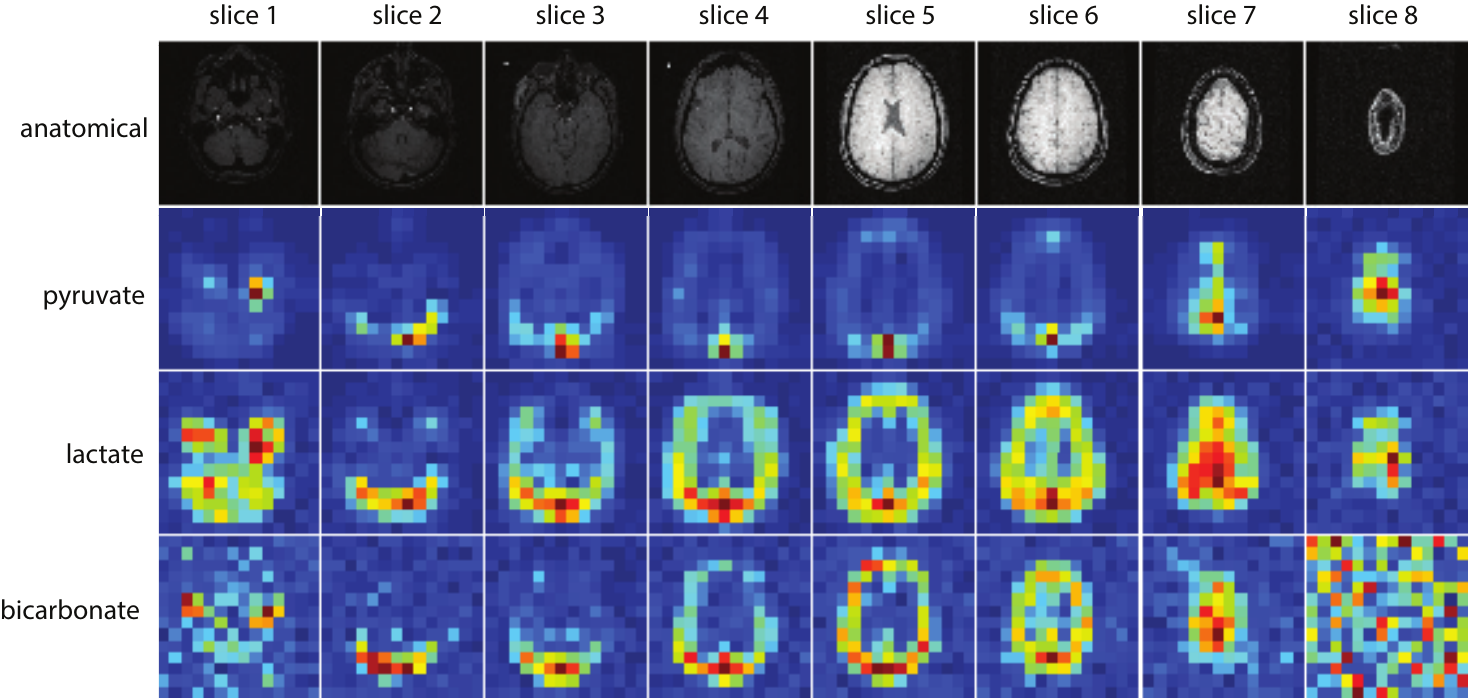}
 \caption
    {
Sum of first twenty time points of multi-slice pyruvate/lactate/bicarbonate dynamic images as described in Fig. 7 and Sup. Fig. S3. Anatomical images are provided as coarse anatomical landmarks. Maximum SNRs of sum-over-time images for pyruvate, lactate and bicarbonate are 600, 58 and 19, respectively.
    }
  \label{fig:human_auc}
 }}
\end{figure}

Results of the hyperpolarized [1-$^{13}$C]pyruvate human experiment using the proposed scheme to image the brain are shown in Fig. \ref{fig:human} and Fig. \ref{fig:human_auc}. The polarization of the injected dose, back-calculated to the time of dissolution, was 41.9\%. The ROI for both bolus tracking and B$_1$ calibration was on the brain tissue near the superior sagittal sinus of slice 5. The frequency spectrum (Fig. \ref{fig:human}c) acquired from slice 5 upon the bolus arrival shows that the measured pyruvate frequency was the same as the frequency calibrated in pre-scan. Fig. \ref{fig:human}b depicts the normalized B$_1$ map which is homogenous as expected for the birdcage transmit coil. The real-time B$_1$ scaling factor (Fig. \ref{fig:human}b) was $\sim$1.04, 4\% difference from the power calibrated in pre-scan. Pyruvate, lactate and bicarbonate sum-over-time images are displayed in Fig. \ref{fig:human_auc}, where maximum SNRs for pyruvate, lactate and bicarbonate are 627, 60 and 20, respectively. Dynamic images can be found in Sup. Fig. S3. 

Experiment recordings are provided on \hl{YouTube} \leftnotepar{\redtext{R1.10\newline R2.9}} (see links in the figure) to demonstrate real-time actions during the experiments. 
\section*{Discussion}

As discussed in the \hl{Introduction} \leftnotepar{\redtext{R2.32}} section, automatic acquisition timing via bolus tracking improves efficient use of hyperpolarization as well as consistency and accuracy of hyperpolarized $^{13}$C imaging. Furthermore, accurate timing is critical for some optimal variable flip angle schemes and metabolic quantification \cite{xing_optimal_2013,Nagashima:2008aa,zhao_gradient-echo_1996,Maidens:2016aa}. The implementation of regional bolus tracking in this work improves the flexibility and accuracy of acquisition timing, and it would be especially useful for imaging tissue with low perfusion rate where injected metabolites would arrive at the target region later relative to major vessels or other tissues. The resolution \hl{of the bolus} \leftnotepar{\redtext{R2.33}} tracking image should be chosen as coarse as possible to improve image SNR but fine enough to separate the desired tracking region (e.g. major vessels from other tissues). An alternative approach is to use 2D spatially selective excitation pulses, but this approach is less robust to signals from off-resonance metabolites.\hl{ In this study, bolus tracking triggered its following sequences either upon bolus arrival or following peak detection.} \leftnotepar{\redtext{RA.1}} This could easily be modified to use other schemes, such as starting metabolic imaging during the bolus or adding an additional delay from the peak. Manually triggering sequences during bolus tracking is also allowed in the designed framework and is useful if the automatic tracking algorithm fails. The bolus tracking used a flip angle of $5^{o}$ and a TR of 1 s. Assuming an injection time of 10s, the expected total hyperpolarized signal loss due to bolus tracking RF pulses would be less than 2\%. 

The naturally low-abundance $^{13}$C signal requires real-time B$_1$ mapping for accurate in vivo measurements of RF power in hyperpolarized $^{13}$C imaging. The acquired B$_1$ map is useful for flip angle correction of images and is crucial for quantification of metabolism (e.g. k$_{PL}$\cite{Sun:2017aa}) as demonstrated in in vivo results (Fig. \ref{fig:liver_coil}). Real-time B$_1$ calibration could achieve accurate flip angle during the scan, which is critical for variable flip angle schemes  \cite{xing_optimal_2013,Nagashima:2008aa,zhao_gradient-echo_1996,Maidens:2016aa} and large flip angle pulses \cite{Schulte:2013aa}, and avoiding unnecessary use of hyperpolarization. The upper limit of the B$_1$ measurement depends on the $K_{BS}$ value of the Bloch-Siegert pulse while the lower limit depends on the SNR. 2D phase unwrapping could be implemented in real time to extend the upper limit of B$_1$ measurement. 
Real-time masking of the acquired B$_1$ map based on signal intensity made the real-time B$_1$ calibration robust when the B$_1$ calibration ROI contained regions with low hyperpolarized signal. The two TRs required for the B1 mapping sequence used a pyruvate flip angle of 10$^{o}$, using only 3\% of the hyperpolarization magnetization. 

Center frequency calibration is crucial for metabolite specific imaging. For singleband spetral-spatial pulses, frequency calibration errors will reduce the flip angle and may also excite undesired resonances.  For spiral or echo-planar imaging (EPI) readouts, frequency calibration errors will lead to off-resonance blurring and shift artifacts.  Both of these factors can undermine the accuracy of Bloch-Siegert B1 mapping and quantification of metabolism. Using a thermal $^{13}$C phantom, which \hl{must} \leftnotepar{\redtext{R2.34}} be placed external to the subject, center frequency calibration doesn't capture B$_0$ within the subject. Although B$_0$ maps can be obtained via proton imaging, in vivo shimming could still be challenging, particularly in the presence of motion. In this study, real-time center frequency was triggered upon bolus arrival, which improved the quality of all the following real-time reconstructed images (Fig. \ref{fig:tramp}e). The 102ms readout duration of the center frequency measurement sequence resulted in a $\sim$10Hz frequency resolution which could be improved with a longer readout duration. A 3$^{o}$ flip angle cost less than 0.2\% hyperpolarization. 

\yellowbox{\partextwidth{The transition time between acquisitions is mainly attributed to real-time image reconstruction and loading sequence waveforms. For animal experiments using single-channel acquisition, the transition times of a bolus tracking acquisition, real-time center frequency calibration, real-time B$_1$ calibration and their following sequences were about 60ms, 60ms and 400ms, respectively. For the human experiment using 32-channel acquisition, a longer transition time ($\sim$800ms) was observed for real-time B$_1$ calibration, whereas transition times for bolus tracking acquisition and real-time center frequency calibration remained similar. The longer transition time found in B$_1$ calibration could be caused by its more complicated reconstruction compared to frequency calibration, and its shorter TR compared to a bolus tracking acquisition which typically used a TR of 1s and allowed real-time reconstruction to be completed during the dead time of a TR. }}\leftnotepar{\redtext{RA.1}}

A $^{13}$C-pyruvate and $^{13}$C-urea co-polarized injection \cite{Wilson:2010aa} would benefit substantially from the proposed methods.  $^{13}$C-urea could be used to perform bolus tracking, real-time center frequency and B$_1$ calibration. This strategy would fully reserve hyperpolarized signal of pyruvate\cite{durst_bolus_2014} and provide higher SNR for B$_1$ mapping by using a larger flip angle. Cardiac and respiratory motion could detrimentally affect bolus tracking and Bloch-Siegert B$_1$ mapping. Cardiac gating could be used \cite{lau_integrated_2012}. In terms of respiratory motion, breathholding is often used in clinical studies. Respiratory gating is not suitable for bolus tracking due to large potential delays but can be applied to perform B$_1$ mapping at the end of exhalation\cite{lau_integrated_2012}, in which case, a short TR (200ms) as used in this study is recommended. A more challenging but robust approach to handle motion is to perform real-time image registration. \leftnotepar{\redtext{RA.1}}\hl{To extend the proposed scheme for volumetric calibration, real-time center frequency and B$_1$ calibration can be integrated with a multislice imaging framework}\cite{lau_integrated_2012}. 
\yellowbox{\partextwidth{
More than one ROI for tracking or B$_1$ calibration could be useful when B$_1$ (e.g. Fig. \ref{fig:liver_coil}b) or acquisition timing (e.g. Fig. \ref{fig:healthy_rat}c) shows variation within the imaging subject.
}}\leftnotepar{\redtext{R2.13}}

\section*{Conclusion}
This work demonstrated an approach that integrates automatic acquisition timing using the regional bolus signal with real-time center frequency calibration, Bloch-Siegert B$_1$ mapping and \hl{real-time RF power compensation based on regional B$_1$ measurements} \leftnotepar{\redtext{R2.2}} as well as dynamic hyperpolarized $^{13}$C imaging. This scheme allows for timing the acquisition based on bolus information of a local region within the imaging plane. \leftnotepar{\redtext{R2.14\newline R2.15}}\hl{Acquiring in vivo B$_1$ maps, performing real-time center frequency calibration and B$_1$ calibration, ensures accurate center frequency and flip angles which are used in the following hyperpolarized $^{13}$C sequences. The proposed scheme was successfully demonstrated for in vivo hyperpolarized [1-$^{13}$C]pyruvate imaging on a clinical 3T scanner.} \hl{Future work will focus on incoporating pyruvate-urea co-polarized injections, volumetric calibration and motion-compensation methods into the proposed scheme to improve efficiency, accuracy, and robustness of hyperpolarized $^{13}$C studies.} \leftnotepar{\redtext{RA.1}}

\subsection*{Acknowledgements}
\leftnotepar{\redtext{RA.1}}\hl{The authors thank Lucas Carvajal, Mark Van Criekinge, James Slater, Mary Mcpolin, Kimberly Okamoto, Hsin-Yu Chen, Daniele Mammoli, Yiran Chen and Peng Cao for their help on the project.} This work was supported by grants R01EB017449, R01EB016741, R01CA183071, and P41EB013598  from the National Institutes of Health.


\newpage
\listoffigures
\newpage
\listoftables

%

\newlength\Colwd
\setlength\Colwd{1.2in}

\begin{table}[h]
\centering  
\begin{tabular}{ | p{1.1\Colwd} | p{1.2\Colwd} |p{0.9\Colwd} |p{0.9\Colwd} |p{0.9\Colwd} |}
\hline
\textbf{Sequence} & \textbf{Parameters} & \textbf{Rat/Birdcage} & \textbf{TRAMP} & \textbf{Rat/Surface} \\ \hline
\multirow{4}{*}{Bolus Tracking} & Slice thickness (mm) & 10 & 20 &10 \\  \cline{2-5}
 & Resolution (mm$^2$) & 2.5 $\times$ 2.5 & 2.5 $\times$ 2.5 & 2.94 $\times$ 2.94\\  \cline{2-5}
 & \multirow{2}{*}{Shared Parameters} & \multicolumn{3}{p{2.7\Colwd}|}{FOV 10cm, FA 5$^o$, TR 1s/0.5s before/after bolus peak, u$_{peak}$ 0.95, c$_{thr}$ 6, n$_{cum}$ 3}  \\ \hline
\multirow{4}{*}{B$_1$ Mapping} & Slice thickness (mm) & 10 & 20 &10 \\  \cline{2-5}
 & Resolution (mm$^2$) & 2.5 $\times$ 2.5 & 2.5 $\times$ 2.5 & 2.94 $\times$ 2.94\\  \cline{2-5}
& \multirow{2}{*}{Shared Parameters} & \multicolumn{3}{p{2.7\Colwd}|}{FOV 10cm, FA 10$^o$, TR 200ms, c$_{B_1}$ 3, \hspace{30pt} desired B$_1$ 0.3G}  \\ \hline 
\multirow{4}{*}{\makecell{Pyruvate/Lactate \\ Dynamic Imaging}} & FOV (cm) & 10 & 8 &10 \\  \cline{2-5}
& Slice thickness (mm) & 10 & 20 &10 \\  \cline{2-5}
 & Resolution (mm$^2$) & 2.5 $\times$ 2.5 & 2.86 $\times$ 2.86 & 2.94 $\times$ 2.94\\  \cline{2-5}
 & \multirow{1}{*}{Shared Parameters} & \multicolumn{3}{p{2.7\Colwd}|}{FA$_{pyr}$ 10$^o$,  FA$_{lac}$ 40$^o$ , Temporal resolution 2s} 
 \\ \hline
 \multirow{2}{*}{\makecell{Frequency \\ Calibration}} & Slice thickness (mm) & NA & 20 &10 \\  \cline{2-5}
 & \multirow{1}{*}{Shared Parameters} & NA & \multicolumn{2}{p{1.8\Colwd}|}{FA 3$^o$, TR 150ms}  \\ \hline
\end{tabular}
\caption{$^{13}$C sequence parameters for animal experiments shown in Fig. \ref{fig:healthy_rat} (`Rat/Birdcage'), Fig. \ref{fig:tramp} (`TRAMP') and Fig. \ref{fig:liver_coil} (`Rat/Surface').}
\label{tab:seq_params}
\end{table}

\begin{table}[h]
\leftnote{\redtext{RA.1}}\yellowbox{\partextwidth{
\centering  
\begin{tabular}{ | p{1.4\Colwd} | p{3.6\Colwd} | }
\hline
\textbf{Sequence} & \textbf{Parameters}\\ \hline
\multirow{2}{*}{Bolus Tracking} & FA 5$^o$, Slice thickness 30mm, Resolution 1.5 $\times$ 1.5 cm$^2$,  \\
& FOV 39cm,  TR 1s,  c$_{thr}$ 4 \\ \hline
\multirow{2}{*}{B$_1$ Mapping} & FA 10$^o$, Slice thickness 30mm, Resolution 1.5 $\times$ 1.5 cm$^2$,  \\ 
& FOV 39cm,  TR 200ms, c$_{B_1}$ 2, desired B$_1$ 0.3G \\ \hline
\multirow{3}{*}{\makecell{Multi-slice Pyruvate/ \\ Lactate/Bicarbonate \\ Dynamic Imaging}} & FA$_{pyr}$ 20$^o$,  FA$_{lac}$ 30$^o$,  FA$_{bicarb}$ 30$^o$, Slice thickness 20mm, \\
& NSlice 8, FOV 39cm, Resolution 1.5 $\times$ 1.5 cm$^2$, TR 125ms, \\
& Temporal resolution 3s\\ \hline
 \multirow{1}{*}{Frequency Calibration} & FA 3$^o$, Slice thickness 20mm, TR 150ms \\ \hline
\end{tabular}
\caption{$^{13}$C sequence parameters for the human study shown in Fig. \ref{fig:human} and Fig. \ref{fig:human_auc}}
\label{tab:human_seq_params}
}}
\end{table}

\newpage
\section*{Supporting Figures}

\begin{enumerate}[label=S\arabic*]
\item{Results of bolus tracking and pyruvate/lactate dynamic imaging of the experiment described in Fig. 4. Fig. 4 shows every other $^{13}$C image in this figure.}
\label{figS:rat_body_supp}

\item{Results of bolus tracking and pyruvate/lactate dynamic imaging of the experiment described in Fig. 5. Fig. 5 shows every other $^{13}$C image in this figure.}
\label{figS:tramp_body_supp}

\item{Sum of first twenty time points of multi-slice pyruvate/lactate/bicarbonate dynamic images as shown in Fig. 8.}
\label{figS:human_supp}
\end{enumerate}


\begin{thebibliography}{10}

\bibitem{Ardenkjaer-Larsen:2003aa}
Ardenkj{\ae}rLarsen~JH, Golman~K, Gram~A, Lerche~MH, Servin~R, Thaning~M,
  Wolber~J.
\newblock Increase of signal-to-noise of more than 10,000 times in liquid state
  NMR.
\newblock Discov Med 2003;3:37--9.

\bibitem{Golman:2006aa}
Golman~K, in~'t Zandt~R, Thaning~M.
\newblock Real-time metabolic imaging.
\newblock Proc Natl Acad Sci U S A 2006;103:11270--5.

\bibitem{Day:2007aa}
Day~SE, Kettunen~MI, Gallagher~FA, Hu~DE, Lerche~M, Wolber~J, Golman~K,
  ArdenkjaerLarsen~JH, Brindle~KM.
\newblock Detecting tumor response to treatment using hyperpolarized 13C
  magnetic resonance imaging and spectroscopy.
\newblock Nat Med 2007;13:1382--7.

\bibitem{Albers:2008aa}
Albers~MJ, Bok~R, Chen~AP, Cunningham~CH, Zierhut~ML, Zhang~VY, Kohler~SJ,
  Tropp~J, Hurd~RE, Yen~YF, Nelson~SJ, Vigneron~DB, Kurhanewicz~J.
\newblock Hyperpolarized 13C lactate, pyruvate, and alanine: noninvasive
  biomarkers for prostate cancer detection and grading.
\newblock Cancer Res 2008;68:8607--15.

\bibitem{Schroeder:2008aa}
Schroeder~MA, Cochlin~LE, Heather~LC, Clarke~K, Radda~GK, Tyler~DJ.
\newblock In vivo assessment of pyruvate dehydrogenase flux in the heart using
  hyperpolarized carbon-13 magnetic resonance.
\newblock Proc Natl Acad Sci U S A 2008;105:12051--6.

\bibitem{Park:2010aa}
Park~I, Larson~PEZ, Zierhut~ML, Hu~S, Bok~R, Ozawa~T, Kurhanewicz~J,
  Vigneron~DB, Vandenberg~SR, James~CD, Nelson~SJ.
\newblock Hyperpolarized 13C magnetic resonance metabolic imaging: application
  to brain tumors.
\newblock Neuro Oncol 2010;12:133--44.

\bibitem{Witney:2010aa}
Witney~TH, Kettunen~MI, Hu~De, Gallagher~FA, Bohndiek~SE, Napolitano~R,
  Brindle~KM.
\newblock Detecting treatment response in a model of human breast
  adenocarcinoma using hyperpolarised [1-$^{13}$C]pyruvate and
  [1,4-$^{13}$C$_{2}$]fumarate.
\newblock Br J Cancer 2010;103:1400--6.

\bibitem{Rider:2013aa}
Rider~OJ, Tyler~DJ.
\newblock Clinical implications of cardiac hyperpolarized magnetic resonance
  imaging.
\newblock J Cardiovasc Magn Reson 2013;15:93.

\bibitem{Darpolor:2011aa}
Darpolor~MM, Yen~YF, Chua~MS, Xing~L, ClarkeKatzenberg~RH, Shi~W, Mayer~D,
  Josan~S, Hurd~RE, Pfefferbaum~A, Senadheera~L, So~S, Hofmann~LV, Glazer~GM,
  Spielman~DM.
\newblock In vivo MRSI of hyperpolarized [1-$^{13}$C]pyruvate metabolism in rat
  hepatocellular carcinoma.
\newblock NMR Biomed 2011;24:506--13.

\bibitem{Kurhanewicz:2011aa}
Kurhanewicz~J, Vigneron~DB, Brindle~K, Chekmenev~EY, Comment~A, Cunningham~CH,
  Deberardinis~RJ, Green~GG, Leach~MO, Rajan~SS, Rizi~RR, Ross~BD, Warren~WS,
  Malloy~CR.
\newblock Analysis of cancer metabolism by imaging hyperpolarized nuclei:
  prospects for translation to clinical research.
\newblock Neoplasia 2011;13:81--97.

\bibitem{Nelson:2013aa}
Nelson~SJ, Kurhanewicz~J, Vigneron~DB, Larson~PEZ, Harzstark~AL, Ferrone~M, van
  Criekinge~M, Chang~JW, Bok~R, Park~I, Reed~G, Carvajal~L, Small~EJ,
  Munster~P, Weinberg~VK, ArdenkjaerLarsen~JH, Chen~AP, Hurd~RE,
  Odegardstuen~LI, Robb~FJ, Tropp~J, Murray~JA.
\newblock Metabolic imaging of patients with prostate cancer using
  hyperpolarized [1-$^{13}$C]pyruvate.
\newblock Sci Transl Med 2013;5:198ra108.

\bibitem{Cunningham:2016aa}
Cunningham~CH, Lau~JYC, Chen~AP, Geraghty~BJ, Perks~WJ, Roifman~I, Wright~GA,
  Connelly~KA.
\newblock Hyperpolarized $^{13}$C Metabolic MRI of the Human Heart: Initial
  Experience.
\newblock Circ Res 2016;119:1177--1182.

\bibitem{Kazan:2013aa}
Kazan~SM, Reynolds~S, Kennerley~A, Wholey~E, Bluff~JE, Berwick~J,
  Cunningham~VJ, Paley~MN, Tozer~GM.
\newblock Kinetic modeling of hyperpolarized $^{13}$C pyruvate metabolism in
  tumors using a measured arterial input function.
\newblock Magn Reson Med 2013;70:943--53.

\bibitem{foo_automated_1997}
Foo~TK, Saranathan~M, Prince~MR, Chenevert~TL.
\newblock Automated detection of bolus arrival and initiation of data
  acquisition in fast, three-dimensional, gadolinium-enhanced MR angiography.
\newblock Radiology 1997;203:275--80.

\bibitem{xing_optimal_2013}
Xing~Y, Reed~GD, Pauly~JM, Kerr~AB, Larson~PEZ.
\newblock Optimal variable flip angle schemes for dynamic acquisition of
  exchanging hyperpolarized substrates.
\newblock J Magn Reson 2013;234:75--81.

\bibitem{Nagashima:2008aa}
Nagashima~K.
\newblock Optimum pulse flip angles for multi-scan acquisition of
  hyperpolarized NMR and MRI.
\newblock J Magn Reson 2008;190:183--8.

\bibitem{zhao_gradient-echo_1996}
Zhao~L, Mulkern~R, Tseng~CH, Williamson~D, Patz~S, Kraft~R, Walsworth~RL,
  Jolesz~FA, Albert~MS.
\newblock Gradient-echo imaging considerations for hyperpolarized $^{129}$Xe
  MR.
\newblock J Magn Reson B 1996;113:179--83.

\bibitem{Maidens:2016aa}
Maidens~J, Gordon~JW, Arcak~M, Larson~PEZ.
\newblock Optimizing flip angles for metabolic rate estimation in
  hyperpolarized carbon-13 MRI.
\newblock IEEE Trans Med Imaging 2016;35:2403--12.

\bibitem{durst_bolus_2014}
Durst~M, Koellisch~U, Gringeri~C, Janich~MA, Rancan~G, Frank~A, Wiesinger~F,
  Menzel~MI, Haase~A, Schulte~RF.
\newblock Bolus tracking for improved metabolic imaging of hyperpolarised
  compounds.
\newblock J Magn Reson 2014;243:40--6.

\bibitem{yen_imaging_2009}
Yen~YF, Kohler~SJ, Chen~AP, Tropp~J, Bok~R, Wolber~J, Albers~MJ, Gram~KA,
  Zierhut~ML, Park~I, Zhang~V, Hu~S, Nelson~SJ, Vigneron~DB, Kurhanewicz~J,
  Dirven~HAAM, Hurd~RE.
\newblock Imaging considerations for in vivo $^{13}$C metabolic mapping using
  hyperpolarized $^{13}$C-pyruvate.
\newblock Magn Reson Med 2009;62:1--10.

\bibitem{gillies_causes_1999}
Gillies~RJ, Schornack~PA, Secomb~TW, Raghunand~N.
\newblock Causes and effects of heterogeneous perfusion in tumors.
\newblock Neoplasia 1999;1:197--207.

\bibitem{Sun:2017aa}
Sun~CY, Walker~CM, Michel~KA, Venkatesan~AM, Lai~SY, Bankson~JA.
\newblock Influence of parameter accuracy on pharmacokinetic analysis of
  hyperpolarized pyruvate.
\newblock Magn Reson Med 2018;79:3239--48.

\bibitem{lau_integrated_2012}
Lau~AZ, Chen~AP, Cunningham~CH.
\newblock Integrated {Bloch}-{Siegert} {B}1 mapping and multislice imaging of
  hyperpolarized $^{13}$C pyruvate and bicarbonate in the heart.
\newblock Magn Reson Med 2012;67:62--71.

\bibitem{schulte_transmit_2011}
Schulte~RF, Sacolick~L, Deppe~MH, Janich~MA, Schwaiger~M, Wild~JM, Wiesinger~F.
\newblock Transmit gain calibration for nonproton MR using the Bloch-Siegert
  shift.
\newblock NMR Biomed 2011;24:1068--72.

\bibitem{Gudbjartsson:1995aa}
Gudbjartsson~H, Patz~S.
\newblock The {Rician} distribution of noisy {MRI} data.
\newblock Magn Reson Med 1995;34:910--4.

\bibitem{sacolick_b_2010}
Sacolick~LI, Wiesinger~F, Hancu~I, Vogel~MW.
\newblock B1 mapping by Bloch-Siegert shift.
\newblock Magn Reson Med 2010;63:1315--22.

\bibitem{TOWERS1991239}
Towers~D, Judge~T, Bryanston{-}Cross~P.
\newblock Automatic interferogram analysis techniques applied to
  quasi-heterodyne holography and ESPI.
\newblock Opt Lasers Eng 1991;14:239--81.

\bibitem{harrison_comparison_2012}
Harrison~C, Yang~C, Jindal~A, Deberardinis~RJ, Hooshyar~MA, Merritt~M,
  Sherry~AD, Malloy~CR.
\newblock Comparison of kinetic models for analysis of pyruvate-to-lactate
  exchange by hyperpolarized 13C {NMR}.
\newblock NMR Biomed 2012;25:1286--94.

\bibitem{mareyam_2017}
Mareyam~A, Carvajal~A, Xu~D, Gordon~J, Park~I, Vigneron~DB, Nelson~SJ,
  Stockmann~JP, Keil~B, Wald~LL.
\newblock 31-Channel brain array for hyperpolarized $^{13}$C imaging at 3T.
\newblock Proc. Intl. Soc. Mag. Reson. Med. 25 2017:1225.

\bibitem{Gordon:2018aa}
Gordon~JW, Hansen~RB, Shin~PJ, Feng~Y, Vigneron~DB, Larson~PEZ.
\newblock 3D hyperpolarized C-13 EPI with calibrationless parallel imaging.
\newblock J Magn Reson 2018;289:92--99.

\bibitem{zhu_2018}
Zhu~Z, Zhu~X, Ohliger~M, Cao~P, Tang~S, Gordon~JW, Aggarwal~R, Bok~R,
  Kurhanewicz~J, Munster~P, Larson~PEZ, Vigneron~DB.
\newblock Coil Combination Methods using Multi-channel Hyperpolarized $^{13}$C
  Clinical Spectroscopic Imaging Data.
\newblock Proc. Intl. Soc. Mag. Reson. Med. 26 2018:8706.

\bibitem{chen_2017}
Chen~HY, Larson~PEZ, Bok~RA, von Morze~C, Sriram~R, DelosSantos~R,
  DelosSantos~J, Gordon~JW, Bahrami~N, Ferrone~M, Kurhanewicz~J, Vigneron~DB.
\newblock Assessing Prostate Cancer Aggressiveness with Hyperpolarized
  Dual-Agent 3D Dynamic Imaging of Metabolism and Perfusion.
\newblock Cancer Res 2017;77:3207--16.

\bibitem{Schulte:2013aa}
Schulte~RF, Sperl~JI, Weidl~E, Menzel~MI, Janich~MA, Khegai~O, Durst~M,
  ArdenkjaerLarsen~JH, Glaser~SJ, Haase~A, Schwaiger~M, Wiesinger~F.
\newblock Saturation-recovery metabolic-exchange rate imaging with
  hyperpolarized [1-$^{13}$C] pyruvate using spectral-spatial excitation.
\newblock Magn Reson Med 2013;69:1209--16.

\bibitem{Wilson:2010aa}
Wilson~DM, Keshari~KR, Larson~PEZ, Chen~AP, Hu~S, VanCriekinge~M, Bok~R,
  Nelson~SJ, Macdonald~JM, Vigneron~DB, Kurhanewicz~J.
\newblock Multi-compound polarization by DNP allows simultaneous assessment of
  multiple enzymatic activities in vivo.
\newblock J Magn Reson 2010;205:141--7.

\end{thebibliography}


\newpage

\end{document}


\newcommand{\comment}[1]{\todo[inline]{#1}}
\subsection*{Supporting Figures}
\renewcommand{\figurename}{Supporting Figure}
\renewcommand{\thefigure}{S\arabic{figure}}
\newcommand\rightnote{\normalmarginpar\marginnote}
\newcommand\leftnote{\reversemarginpar\marginnote}
\newcommand\leftnotepar{\reversemarginpar\marginpar}
\newcommand\rightnotepar{\normalmarginpar\marginpar}
\newcommand\partextwidth{\parbox{1\textwidth}}
\newcommand\yellowbox{\colorbox{yellow}}
\newcommand\redtext{\color{red}}

\renewcommand\hl{\textnormal}
\renewcommand\yellowbox{\textnormal}
\renewcommand\partextwidth{\textnormal}
\renewcommand\leftnote{\comment}
\renewcommand\rightnote{\comment}
\renewcommand\rightnotepar{\comment}
\renewcommand\leftnotepar{\comment}

\begin{figure} 

\centering  
\includegraphics[width=6.5in]{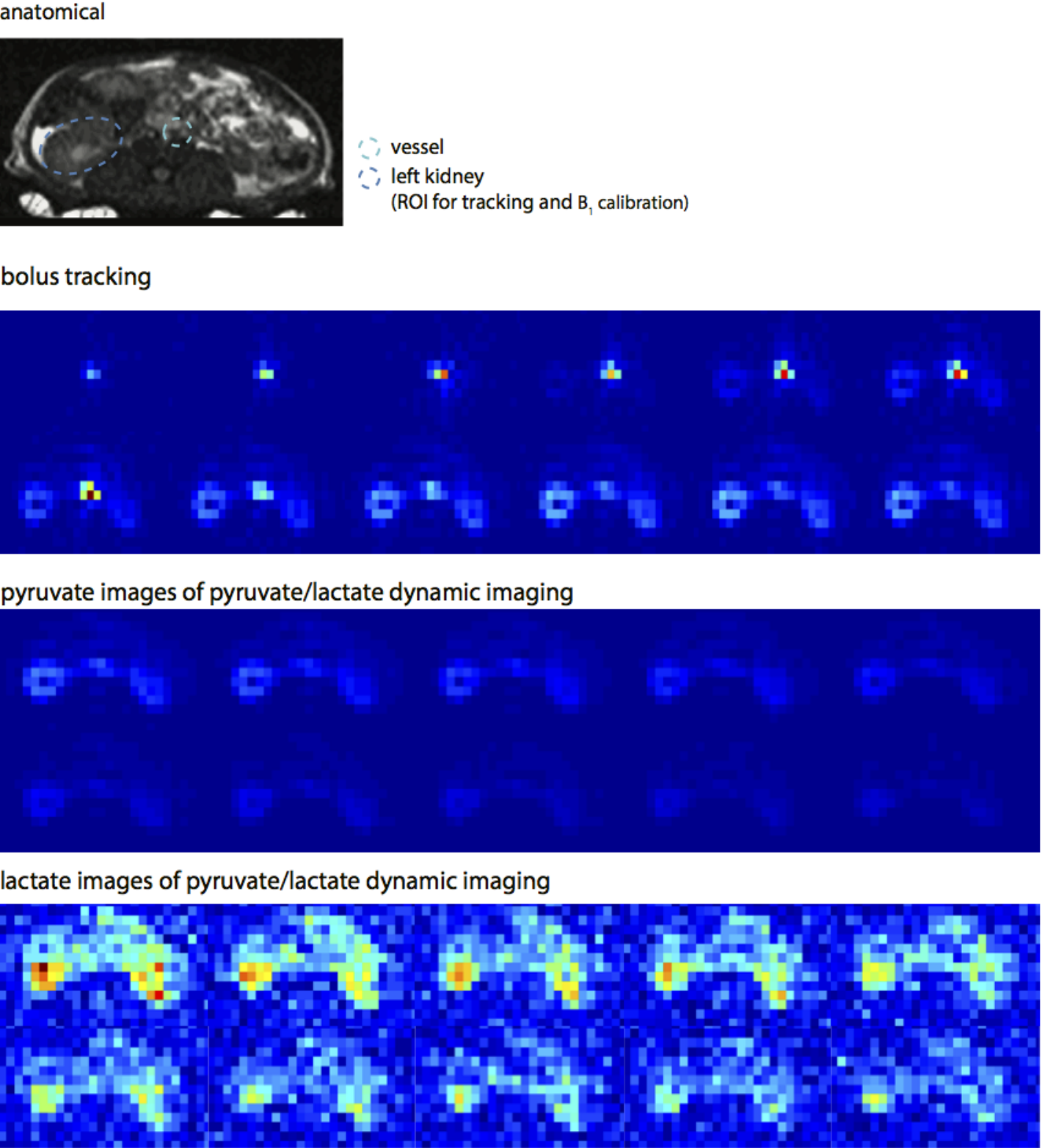}
 \caption
    {
Results of bolus tracking and pyruvate/lactate dynamic imaging of the experiment described in Fig. 4. Fig. 4 shows every other $^{13}$C image in this figure.
    }
  \label{suppl:rat_body_supp}
\end{figure}

\begin{figure} 
\centering  
\includegraphics[width=6in]{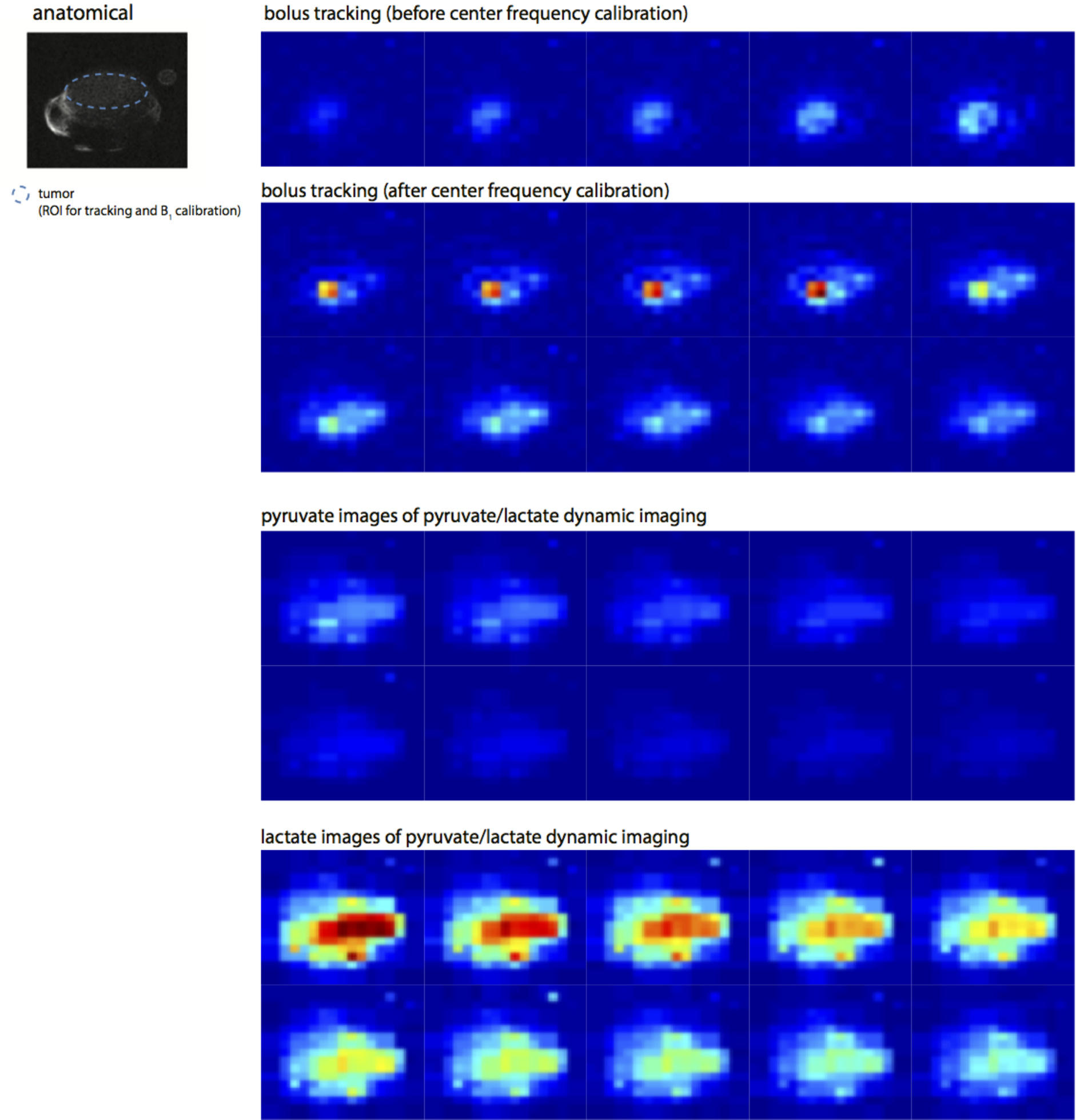}
 \caption
    {
Results of bolus tracking and pyruvate/lactate dynamic imaging of the experiment described in Fig. 5. Fig. 5 shows every other $^{13}$C image in this figure.
    }
  \label{suppl:tramp_body_supp}
\end{figure}

\begin{figure} 
\centering  

 \leftnote{\redtext{RA.1}}
 \yellowbox{\partextwidth{
  \includegraphics[width=1\textwidth]{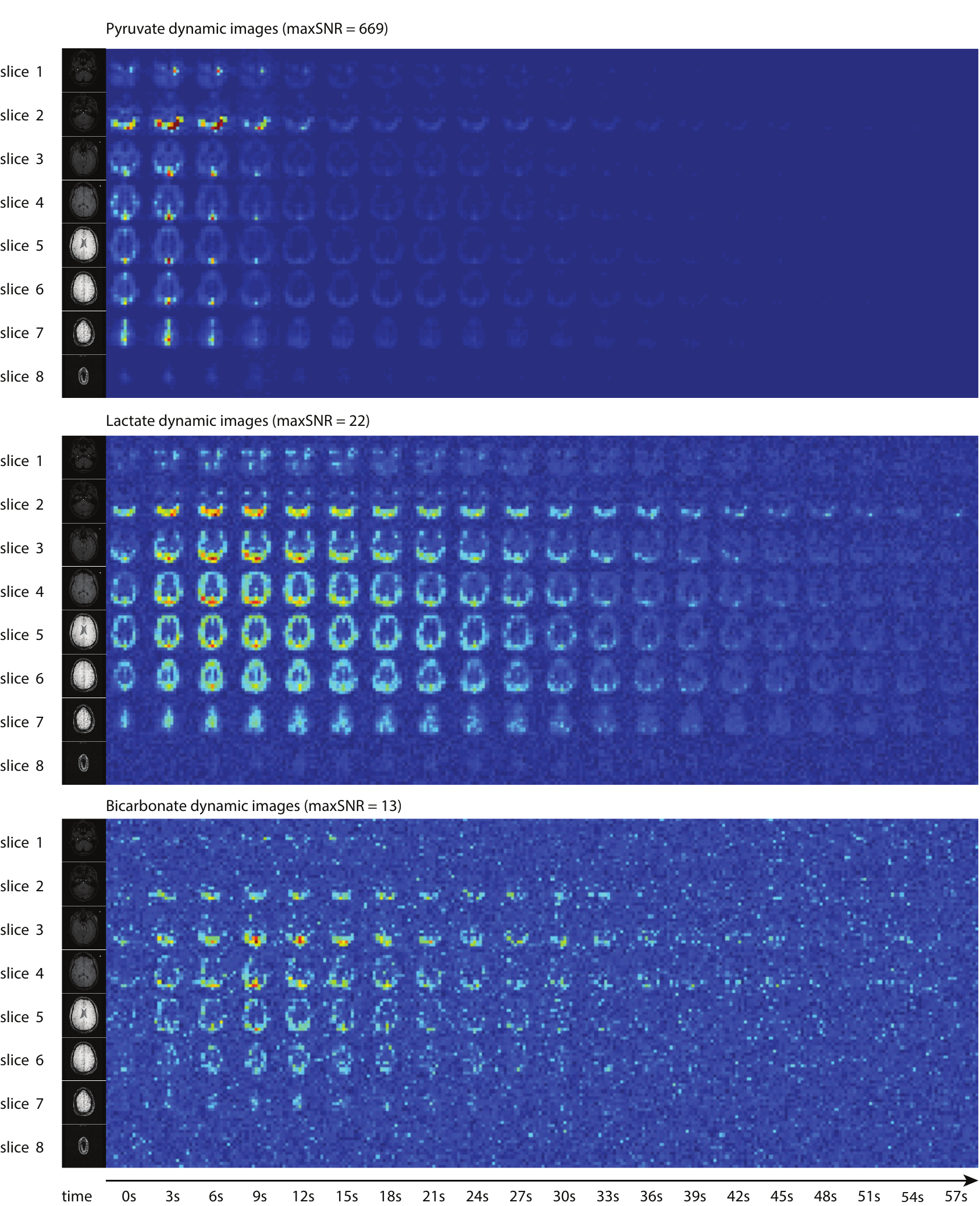}
    \caption
    {
The first twenty time points of multi-slice dynamic images for the hyperpolarized [1-$^{13}$C]pyruvate human study as described in Fig. 7. 
    }
  \label{suppl:human_all_dynamics}    
}}
    
\end{figure}